\documentclass{emulateapj}
\usepackage{hyperref}

\usepackage[dvipsnames]{xcolor}
\colorlet{mylinkcolor}{Maroon}
\colorlet{mycitecolor}{MidnightBlue}
\colorlet{myurlcolor}{MidnightBlue}
\hypersetup{ 
	colorlinks   = true,
	citecolor    = mycitecolor,
	linkcolor    = mylinkcolor,
	urlcolor	= myurlcolor
}

\usepackage{etoolbox}
\makeatletter
\patchcmd{\BR@backref}{\newblock}{\newblock[}{}{}
\patchcmd{\BR@backref}{\par}{]\par}{}{}
\makeatother
\usepackage[all]{hypcap}

\usepackage{graphicx}
\usepackage{natbib}
\usepackage{amsmath,amssymb}
\usepackage{xcolor}
\usepackage[flushleft]{threeparttable}
\usepackage{booktabs}

\usepackage{CJKutf8}
\usepackage{ucs}
\usepackage[encapsulated]{CJK}
\newcommand{\myfont}{bsmi} 

\submitted{ApJ in press}

\lefthead{Chen et al.}
\righthead{A spatially-resolved multi-wavelength study of ALESS67.1}

\begin{document}


\title{A spatially-resolved study of cold dust, molecular gas, H{\sc ii} regions and stars in a $\MakeLowercase{z}=2.12$ submillimeter galaxy ALESS67.1}

\begin{CJK}{UTF8}{\myfont} 
\author{Chian-Chou Chen (陳建州)\altaffilmark{1,2}, J. A. Hodge\altaffilmark{3}, Ian Smail\altaffilmark{2}, A. M. Swinbank\altaffilmark{2}, Fabian Walter\altaffilmark{4},  J. M. Simpson\altaffilmark{5}, Gabriela Calistro Rivera\altaffilmark{3}, F. Bertoldi\altaffilmark{6}, W. N. Brandt\altaffilmark{7,8,9}, S. C. Chapman\altaffilmark{10}, Elisabete da Cunha\altaffilmark{11}, H. Dannerbauer\altaffilmark{12,13}, C. De Breuck\altaffilmark{1}, C. M. Harrison\altaffilmark{1}, R. J. Ivison\altaffilmark{1,14}, A. Karim\altaffilmark{6}, K. K. Knudsen\altaffilmark{15}, J. L. Wardlow\altaffilmark{2}, A. Wei{\ss}\altaffilmark{16}, P. P. van der Werf\altaffilmark{3}}

\email{ccchen@eso.org}
\altaffiltext{1}{European Southern Observatory, Karl Schwarzschild Strasse 2, Garching, Germany}
\altaffiltext{2}{Centre for Extragalactic Astronomy, Department of Physics, Durham University, South Road, Durham DH1 3LE, UK}
\altaffiltext{3}{Leiden Observatory, Leiden University, P.O. Box 9513, 2300 RA Leiden, the Netherlands}
\altaffiltext{4}{Max–Planck Institut f\"{u}r Astronomie, K\"{o}nigstuhl 17, 69117 Heidelberg, Germany}
\altaffiltext{5}{Academia Sinica Institute of Astronomy and Astrophysics, No. 1, Sec. 4, Roosevelt Rd., Taipei 10617, Taiwan}
\altaffiltext{6}{Argelander–Institute  of  Astronomy,  Bonn  University,  Auf dem H\"{u}gel 71, D–53121 Bonn, Germany}
\altaffiltext{7}{Department of Astronomy \& Astrophysics, 525 Davey Lab, Pennsylvania State University, University Park, PA 16802, USA.}
\altaffiltext{8}{Institute for Gravitation and the Cosmos, Pennsylvania State University, University Park, PA 16802, USA.}
\altaffiltext{9}{Department of Physics, 104 Davey Lab, The Pennsylvania State University, University Park, PA 16802, USA}
\altaffiltext{10}{Department of Physics and Atmospheric Science, Dalhousie University, Halifax, NS B3H 3J5, Canada.}
\altaffiltext{11}{The Australian National University, Mt Stromlo Observatory, Cotter Rd, Weston Creek, ACT 2611, Australia}
\altaffiltext{12}{Instituto de Astrofísica de Canarias (IAC), E-38205 La Laguna, Tenerife, Spain}
\altaffiltext{13}{Universidad de La Laguna, Dpto. Astrofísica, E-38206 La Laguna, Tenerife, Spain}
\altaffiltext{14}{Institute for Astronomy, University of Edinburgh, Royal Observatory, Blackford Hill, Edinburgh EH9 3HJ, UK.}
\altaffiltext{15}{Department of Earth and Space Sciences, Chalmers University of Technology, Onsala Space Observatory, 439 92 Onsala, Sweden}
\altaffiltext{16}{Max-Planck-Institut f\"{u}r Radioastronomie, Auf dem H\"{u}gel 69, D-53121 Bonn, Germany}

\begin{abstract}
We present detailed studies of a $z=2.12$ submillimeter galaxy, ALESS67.1, using sub-arcsecond resolution ALMA, AO-aided VLT/SINFONI, and {\it HST}/CANDELS data to investigate the kinematics and spatial distributions of dust emission (870\,$\mu$m continuum), $^{12}$CO($J$=3-2), strong optical emission lines, and visible stars. Dynamical modelling of the optical emission lines suggests that ALESS67.1 is not a pure rotating disk but a merger, consistent with the apparent tidal features revealed in the {\it HST} imaging. 
Our sub-arcsecond resolution dataset allow us to measure half-light radii for all the tracers, and we find a factor of 4--6 smaller sizes in dust continuum compared to all the other tracers, including $^{12}$CO, and  UV and H$\alpha$ emission is significantly offset from the dust continuum. The spatial mismatch between UV continuum and the cold dust and gas reservoir supports the explanation that geometrical effects are responsible for the offset of dusty galaxy on the IRX-$\beta$ diagram. Using a dynamical method we derive an $\alpha_{\rm CO}=1.8\pm1.0$, consistent with other SMGs that also have resolved CO and dust measurements. Assuming a single $\alpha_{\rm CO}$ value we also derive resolved gas and star-formation rate surface densities, and find that the core region of the galaxy ($\lesssim5$\,kpc) follows the trend of mergers on the Schmidt-Kennicutt relationship, whereas the outskirts ($\gtrsim5$\,kpc) lie on the locus of normal star-forming galaxies, suggesting different star-formation efficiencies within one galaxy. Our results caution against using single size or morphology for different tracers of the star-formation activity and gas content of galaxies, and therefore argue the need to use spatially-resolved, multi-wavelength observations to interpret the properties of SMGs, and perhaps even for $z>1$ galaxies in general.
 
\end{abstract}


\section{Introduction}\label{sec:intro}
Recent technical advances in instruments now allow astronomers to conduct spatially-resolved, multi-wavelength observations of astronomical sources. This is particularly important as observations in different wavelengths probe different physical processes, and only by combining the data across many wavelengths is it possible to put together a complete picture of galaxy formation and evolution and draw an unbiased conclusion.

The importance of spatially-resolved, multi-wavelength observations is well illustrated in the local Universe. Surveys of nearby galaxies in a variety of wavebands have offered great legacy value, including census of star-forming regions and young stars in the ultraviolet (UV; \citealt{Gil-de-Paz:2007aa}) and optical \citep{Gunn:2006aa}, dust distributions in the infrared (IR; \citealt{Kennicutt:2003aa, Kennicutt:2011aa}), as well as molecular gas traced in the millimeter by CO \citep{Leroy:2009aa} and at radio wavelength for atomic hydrogen \citep{Walter:2008aa}. However, it is only by combining these surveys that fundamental insights into galaxy formation, such as the Schmidt-Kennicutt relationship, is revealed (e.g., \citealt{Kennicutt:1998p5718,Leroy:2008aa, Sandstrom:2013aa}). 

At high redshifts, however, where observations suffer from cosmological dimming and typically smaller galaxy sizes, obtaining sensitive multi-wavelength datasets on a common galaxy sample becomes difficult. 
This is particularly true for dust-obscured populations such as submillimeter galaxies (SMGs; \citealt{Smail:1997p6820, Barger:1998p13566, Hughes:1998p9666}), or more generally the class of dusty star-forming galaxies (DSFGs; \citealt{Casey:2014aa}). 

SMGs are submillimeter-bright dusty galaxies which are shown to be forming stars at some of the highest rates known, with star-formation rates (SFRs) up to $\sim$1000\,M$_\odot$\,yr$^{-1}$ (e.g., \citealt{Barger:2012lr, Swinbank:2014aa}). For 850\,$\mu$m-selected SMGs they are found to be most prevalent at $z\sim2-3$ (e.g., \citealt{Chapman:2005p5778, Wardlow:2011qy, Simpson:2014aa, Chen:2016aa}), corresponding to the peak of the cosmic SFR density \citep{Madau:2014aa}, and they appear to be some of the most massive galaxies existing during that epoch (e.g., \citealt{Barger:2014aa}). Therefore since their discovery, SMGs have provided an ideal laboratory for testing the physical conditions in which the extreme star formation occurs, both theoretically (e.g., \citealt{Baugh:2005p14519,Dave:2010kx,Hayward:2013lr,Cowley:2015aa}) and observationally (e.g., \citealt{Swinbank:2006aa,Bouche:2007aa,Bothwell:2010p10377,Alaghband-Zadeh:2012aa,Sharon:2013aa,Rawle:2014aa,Hodge:2015aa}). 

Among the available observational tests, measurements of galaxy dynamics through ionized or molecular gas and the spatial distribution of dust and stars have the most distinguishing power between models (e.g., \citealt{Narayanan:2010aa,Bournaud:2014aa}). However, obtaining these data is also the most difficult due to the requirement of high ($\sim0\farcs1$) spatial resolution. For $z\sim2$ SMGs in the blank field this can only be achieved with near-infrared integral field unit (IFU) observations aided with adaptive optics (AO) for redshifted optical emission lines such as H$\alpha$ (e.g., \citealt{Alaghband-Zadeh:2012aa}), interferometers to obtain resolved far-IR/(sub-)millimeter continuum or CO (e.g., \citealt{Younger:2008rt}), and space-based observatories such as the {\it Hubble Space Telescope} ({\it HST}; e.g., \citealt{Swinbank:2010aa}) to provide diffraction-limited UV-to-NIR imaging of the stellar continuum. The rarity of SMGs means that detailed SMG studies to date have either focused on the UV/optical/NIR (e.g., \citealt{Menendez-Delmestre:2013aa,Chen:2015aa}) or the FIR/submillimeter (e.g., \citealt{Danielson:2011aa,Spilker:2014aa,ALMA-Partnership:2015aa}).

The need to combine UV/optical/NIR and FIR/submillimeter imaging on individual sources is driven by the significant differences sometimes found when comparing results from the two types of study. First, studies of H$\alpha$ dynamics have found that SMGs are mostly dispersion-dominated systems and are consistent with them being mergers \citep{Alaghband-Zadeh:2012aa,Menendez-Delmestre:2013aa,Olivares:2016aa}, whereas the kinematics of CO and [C\,{\sc ii}] on some of the other samples of SMGs have been shown that they resemble the structures of rotating disks (e.g., \citealt{Hodge:2012fk,De-Breuck:2014aa}). While part of this could be the different relaxation time scale between gas and H{\sc ii} regions (e.g., \citealt{Hopkins:2013aa}), it could also be that the kinematic of CO and H$\alpha$ are in fact consistent with each other once measured on the same galaxies, and the different results are genuine variations simply due to small numbers of sources in both types of study. 

Such resolved studies would help answer various open questions about SMGs. For example, by compiling a sample of $z<3.5$ DSFGs that have rest-frame UV coverage, \citet{Casey:2014ab} have found significantly bluer UV continuum slopes ($\beta$) than the local star-forming galaxy (SFG) samples given a fixed IR-to-UV luminosity ratio (IRX). Casey et al. had argued that the geometrical effects in which a mismatch between the bulk of IR and UV emissions, which is also observed in the local ultra-luminous infrared galaxies (ULIRGs; \citealt{Sanders:1996p6419}), could be one of the most important factors that causes the deviation of DSFGs from the nominal IRX-$\beta$ relationship. By combining high resolution imaging of the optical and dust emission we can test these hypotheses.

Similarly, by modelling the UV-to-NIR spectral energy distributions (SEDs) it has been found that the dust extinction against the NIR-detectable stellar continuum of SMGs is typically of $A_{\rm V}\sim1-3$ (e.g., \citealt{da-Cunha:2015aa}), in contrast with the estimates ($A_{\rm V}\sim500$) based on the column density of dust where the size of the dusty regions is available \citep{Simpson:2017ab}. Although these two studies were conducted using different SMG samples, Simpson et al. argued that the relative compact sizes and distributions of dust with respect to the UV-to-NIR continuum could be the main cause for the discrepancy, simply because the flux-weighted SED modelling based on UV-to-NIR photometry is not reflecting the majority of the dust extinction that is coming from a more compact and very dense and dusty region. These examples illustrate that having spatially-resolved panchromatic data with both photometry and spectroscopy on the same galaxies is the key to make further progress on these issues.

Here we present such a study of the $z=2.12$ SMG ALESS67.1, where we have collected sub-arcsecond UV-to-NIR continuum from the {\it HST}, NIR IFU from the AO-aided SINFONI observations, and 870\,$\mu$m continuum and $^{12}$CO($J$=3-2) from the Atacama Large Millimeter/submillimeter Array (ALMA). ALESS67.1 is part of the ALESS sample \citep{Karim:2013fk,Hodge:2013lr}, a Cycle 0 ALMA survey targeting a flux-limited sample of 126 submillimeter sources detected by a LABOCA \citep{Siringo:2009rt} 870\,$\mu$m survey in the Extended {\it Chandra} Deep Field South (ECDFS) field (LESS survey; \citealt{Weis:2009qy}). 

ALESS67.1 is one of the few SMGs so far that is covered by all the necessary follow-up observations, and it is representative of the ALESS sample; ALESS67.1 has a spectroscopic redshift at $z_{\rm spec}=2.1230$ \citep{Danielson:2017aa} with a SFR of $\sim500$\,M$_\odot$\,yr$^{-1}$ \citep{Swinbank:2014aa,da-Cunha:2015aa} and a stellar mass of $\sim2\times10^{11}$\,M$_\odot$ \citep{Simpson:2014aa,da-Cunha:2015aa}. ALESS67.1 appears to be a merger remnant in the {\it HST} imaging \citep{Chen:2015aa}, and it is detected by {\it Chandra} in the 0.5--2\,keV X-ray band \citep{Wang:2013aa}. However because of its relatively low X-ray luminosity ($L_{\rm 0.5-8keV}$=$3\times10^{42}$\,erg\,s$^{-1}$), Wang et al. concluded that the X-ray luminosity might be contributed by both star-formation and AGN, which is consistent with the optical line ratios \citep{Danielson:2017aa}, indicating that ALESS67.1 lies in the composite region of the BPT diagram. Here we include in our analyses the high-resolution ALMA 870\,$\mu$m continuum observations, $^{12}$CO($J$=3-2), and AO-aided SINFONI. The data reduction and analyses of these data are presented in \autoref{sec:obs} and our results are in \autoref{sec:analyses}. We discuss in \autoref{sec:discussion} regarding the kinematics of CO and H$\alpha$, CO-to-H$_2$ conversion factor, the size contrast between dust and other tracers and its implication on the IRX-$\beta$ relationship and the Schmidt-Kennicutt relationship. Finally our conclusions are given in \autoref{sec:summary}.

In this paper we assume the {\it Planck} cosmology: H$_0 =$\,67.77\,km\,s$^{-1}$ Mpc$^{-1}$, $\Omega_M = $\,0.31, and $\Omega_\Lambda =$\,0.69 \citep{Planck-Collaboration:2014aa}. We also assume a Chabrier initial mass function \citep{Chabrier:2003aa}.

\section{Observations and data reduction}\label{sec:obs}
\subsection{ALMA 870\,$\mu$m continuum}
The ALMA Band 7 data were taken on the 11$^{\rm th}$ August 2015, as part of a Cycle 1 project \#2012.1.00307.S (PI: J. Hodge), which targeted 19 SMGs from the Cycle 0 ALESS survey (Hodge et al. 2013). For a full detail description of the project please refer to \citet{Hodge:2016aa}.

As in the Cycle 0 ALESS program, the Band 7 data were centered on 344\,GHz ($\sim$870\,$\mu$m). We used the ``single continuum'' spectral mode, with $4\times128$ dual polarization channels over the 8\,GHz bandwidth. The primary beam of the ALMA observations is 17$\farcs$4 at full-width-half-maximum (FWHM).

The ALMA data were obtained using 46 antennas in an extended configuration (C32-6; maximum baseline of $\sim$1.6\,km). The bandpass, phase, and flux calibrators were J0522--3627, J0348--2749, and J0334--401, respectively, and the total integration time was approximately eight minutes. The data were taken under good phase stability/weather conditions, with a medium PWV at zenith of $\sim$0.7\,mm.

The {\it uv}--data were inverse Fourier--transformed using natural weighting to produce the dirty continuum image, which was later deconvolved with a synthesized beam (i.e., the dirty beam) using the {\sc clean} algorithm. The image is gridded to a pixel scale of 0$\farcs$02 and a size of  20$\farcs$48 (1024 pixels) per side, covering the primary beam of our observations. The FWHM of the synthesized beam is 0$\farcs$18$\times$0$\farcs$15 (1.5$\times$1.3\,kpc at the redshift of ALESS67.1), with a position angle of 64.8$^\circ$. The r.m.s. noise of the dirty map is 1\,$\sigma=0.07$\,mJy beam$^{-1}$.

\subsection{ALMA $^{12}$CO J=3--2}
The ALMA data were taken on the 6$^{\rm th}$ September 2015 as part of project 2013.1.00470.S (PI: J. Hodge), which targets a sample of SMGs to obtain sub-arcsecond resolution CO maps to study the properties of molecular gas (Calistro Rivera et al.\ in preparation). The data were taken in Band 3, with the expected frequency of the redshifted CO($J$=3-2) line ($\nu_{\rm rest}$ = 345.7959899 GHz) covered by the upper side band, and using three additional basebands to observe the continuum. We used the lowest-resolution FDM mode and averaged over eight channels to maintain adequate resolution while keeping the data rate reasonable. The observations were carried out with 36 antennas in an intermediate configuration (maximum baseline 1.6\,km) and the maximum recoverable scale is 4.7$''$. Standard calibration was used and the total integration time on the target was 29 minutes. Data were reduced using {\sc casa} version 4.3.1 and the standard pipeline calibration, with some additional flags applied to address bad antennas or times. Imaging was carried out using {\sc casa} version 4.7.0. The {\it uv}--data were inverse Fourier--transformed using natural weighting to produce both the 3\,mm continuum dirty image and the CO data cube. There is no detection in the 3\,mm continuum, thus putting a  3\,$\sigma$ constraint on the 3\,mm continuum flux of $S_{\rm 3mm} \leq 0.054$\,mJy. The CO cube is gridded into 48MHz per channel ($\sim$130\,km\,s$^{-1}$) and has an average synthesized beam of 0$\farcs$57$\times$0$\farcs$49. The sensitivity of the CO cube is 0.25\,mJy beam$^{-1}$ per 48MHz channel. 

\begin{figure*}
	\begin{center}
		\leavevmode
		\includegraphics[scale=0.4]{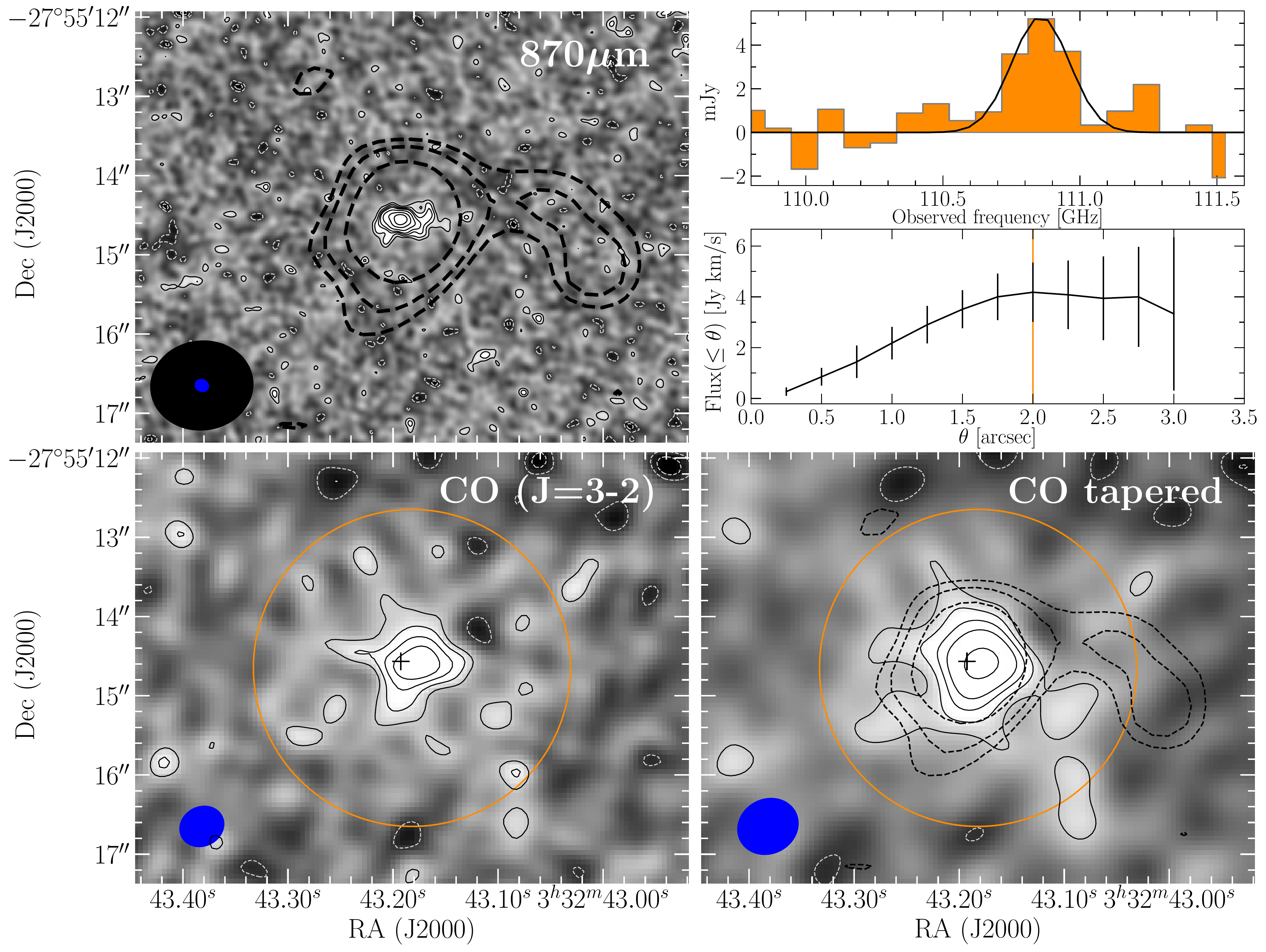}
		\caption{{\it Top-left:} The ALMA $\sim$0$\farcs$2 870\,$\mu$m continuum map with solid contours at levels of [2,3,4,7,10,13]$\times\sigma$. The dotted contours show the detected emission in the $\sim$1$''$ Cycle 0 ALMA data presented in \citet{Hodge:2013lr}, with the levels at [2,3,5]$\times\sigma$. The synthesized beam shapes are shown at the bottom-left corner. {\it Bottom:} FWHM-averaged maps of CO based on the spectrum shown in the top-right panel, which is obtained by summing all the fluxes in the naturally-weighted map (left) within a 2$''$ radius circle (orange circles). The radius of 2$''$ is determined through our curve-of-growth analysis, which is shown below the spectrum. The resolution of the tapered map (right) is $\sim$0$\farcs$7. The solid contours in both maps are [2, 3, 4, 5]$\times \sigma$ and the grey dashed ones are [-3,-2]$\times\sigma$. The dashed contours in the bottom-right panel shows the 870\,$\mu$m continuum from the Cycle 0 ALMA data. The small cross symbols mark the peak location of the 870\,$\mu$m continuum emission.
		}
		\label{cocont}
	\end{center}
\end{figure*}

\subsection{VLT/SINFONI}
AO-assisted, IFU observations of the strong optical lines in ALESS\,67.1 were taken with the SINFONI IFU between 2013 January and October.  At $z=$2.12, the [N\,{\sc ii}]/H$\alpha$ lines are redshifted to $\lambda\sim$\,2.05\,$\mu$m and [O\,{\sc iii}]/H$\beta$ are redshifted to $\lambda\sim1.55$\,$\mu$m so we used the $HK$-band filter and grism which has a spectral resolution of $R$\,=\,$\lambda$\,/\,$\Delta\lambda\sim$\,5000, sufficient to separate H$\alpha$ and the two [N\,{\sc ii}] lines. Since the low-surface brightness continuum emission is spatially extended across $\sim$3$''$ in the \emph{HST} $H$-band imaging, we used the 8\,$\times$\,8$''$ field of view mode of SINFONI.  To achieve high spatial resolution, we employed natural guide star (NGS) AO correction exploiting a nearby bright ($R$\,=\,12.9 mag) star.  Each 1\,hr observation block (OB) was split in to 4\,$\times$\,600\,s exposures, which were dithered by 4$''$, thus always keeping the target in the field-of-view (FOV).  In total, we observed the target for 7.2\,ks.   Data reduction was performed using the {\sc esorex} pipeline, with additional custom routines applied to improve the flat-fielding, sky subtraction and mosaicing of the cubes. The flux and astrometry calibration is calibrated from the Hawk-I $K$-band imaging. The AO-corrected PSF has a mean Stehl ratio of 0.3, ideally corresponding to an angular resolution of $\sim$0.2$''$ FWHM.  

\subsection{{\it HST} optical/NIR imaging}
The optical and near-infrared (NIR) images from the ACS and WFC3 cameras mounted on the {\it Hubble Space Telescope} ({\it HST}) were taken as part of the Cosmic Assembly Near-IR Deep Extragalactic Legacy Survey (CANDELS; \citealt{Grogin:2011fj,Koekemoer:2011aa}). The typical FWHM of the {\it HST} PSF in the optical is $\sim0\farcs1$.

\section{Analysis and Results}\label{sec:analyses}
\subsection{ALMA 870\,$\mu$m continuum}
A prominent source is detected in the central region of the high-resolution dirty map, with a peak flux of 0.72\,mJy\,beam$^{-1}$ (corresponding to 10\,$\sigma$) and a location matching ALESS67.1 from \citet{Hodge:2013lr}. We clean a circular region with 1$''$ radius around the source down to 2\,$\sigma$, and the resulting cleaned image is shown in \autoref{cocont}.  As seen in \autoref{cocont}, two detections were reported at $\sim$1$''$ resolution observations by \citet{Hodge:2013lr}, ALESS67.1/67.2, however in our data we only detect ALESS67.1. We have tried tapering the map to lower spatial resolution in order to test the possibility that the lack of detection is due to extended structures which are resolved out in high-resolution map. While ALESS67.2 remained undetected in the tapered maps, the sensitivity of the tapered map is not as deep as the original Cycle 0 data so the nature of ALESS67.2 remains inconclusive. It is possible that ALESS67.2 is resolved out, or it is also possible that ALESS67.2 is a false detection.

As ALESS67.1 is clearly resolved and it is the sole source detected in the map, to measure the flux and the light profile, we first use the {\sc uvmodelfit} algorithm to model the {\it uv}--data. We find that the best bit (reduced $\chi^2$=0.5) Gaussian profile has an intensity of 3.7$\pm$0.2\,mJy with a FWHM of $0\farcs40\pm0\farcs02\times0\farcs21\pm0\farcs02$, corresponding to a physical half-light radius of $1.7\pm0.1\times0.9\pm0.1$\,kpc. We obtain consistent results if we instead using the {\sc imfit} algorithm or {\sc sextractor} on the cleaned image. 
The measured flux is also consistent but marginally lower than the previous measurement (4.9$\pm$0.7\,mJy) based on the lower resolution ALMA Cycle 0 data \citep{Hodge:2013lr}, and the size is consistent with the parent sample of ALESS SMGs with $0\farcs2$ high-resolution ALMA observations \citep{Hodge:2016aa}.

\subsection{ALMA $^{12}$CO J=3--2}\label{sec:co}

A strong line detection is seen in the dirty 3mm channel maps, and the emission appears resolved. To clean the data cube and extract the spectra, we employ the following iterative procedure; We first derive a weight-averaged map over a best-guess frequency width and choose a center based on the averaged map. We then perform a curve-of-growth analysis where we define a circular aperture centered at the chosen centroid with a radius which encompasses all the line flux. The data cube is then cleaned to 2\,$\sigma$ within this defined aperture and the spectrum is extracted. Next the extracted spectrum is fitted with a Gaussian profile and the frequency width used for obtaining the averaged map is updated to be the FWHM of the spectral fit. This process continues until the solution converges. In the end we find that all the line flux is contained within a circular radius of 2$''$ and the results are not sensitive to the chosen position of the aperture center within a beam area.
 
The results are plotted in \autoref{cocont}, showing a line detection well fitted ($\chi^2 = 1.1$) with a Gaussian profile centered at 110845$\pm$30\,MHz and having a FWHM of 319$\pm$72\,MHz, corresponding to a $^{12}$CO($ J$=3-2) line at a redshift of $z=2.1196\pm0.0009$, with a velocity FWHM of 862$\pm$195\,km\,s$^{-1}$. By integrating the best-fit Gaussian we derive a total line flux of 4.2$\pm$1.2\,Jy\,km\,s$^{-1}$. The errors are estimates from the fit, and they are consistent with the errors derived from a Monte Carlo simulation. We create fake spectra by injecting model profile into spectra extracted from randomly selected regions of the data cube with the same circular aperture used for the detection spectrum. The errors are then obtained from the standard deviations between the fit results and the input model.

Using the standard relation from Solomon \& Vanden Bout (2005), $L^\prime_{\rm CO}=3.25\times10^7 S_{\rm CO}\Delta v\nu_{\rm obs}^{-2}D_{\rm L}^2(1+z)^{-3}$, where $S_{\rm CO}\Delta v$ is the total line flux in Jy\,km\,s$^{-1}$, $\nu_{\rm obs}$ is the observed line frequency in GHz, and $D_{\rm L}$ is the cosmological luminosity distance in Mpc, we calculate a CO luminosity of $L^\prime_{\rm CO(3-2)}=(1.1\pm0.3)\times10^{11}$\,K\,km\,s$^{-1}$\,pc$^2$. 
\citet{Huynh:2017aa} have recently conducted $^{12}$CO(1-0) observations on ALESS67.1 using the Australia Telescope Compact Array (ATCA), and they detect strong $^{12}$CO(1-0) emission with a total flux of $L^\prime_{\rm CO(1-0)}=(9.9\pm1.8)\times10^{10}$\,K\,km\,s$^{-1}$\,pc$^2$. With both the measurements we calculate a $L^\prime_{\rm CO(3-2)}/L^\prime_{\rm CO(1-0)}$ line luminosity ratio of $r_{3,1} = 1.1\pm0.4$, consistent with previous estimates for SMG population \citep{Harris:2010p11118,Ivison:2011aa,Bothwell:2013lp,Sharon:2016aa}. 

The curve-of-growth analysis shown in \autoref{cocont} suggests a CO half-light radius of $\sim1''$. To measure the size we employ both image-based and {\it uv}-based analyses. We first conduct {\sc imfit} on the averaged map and find that the best-fit two-dimensional (2D) Gaussian profile has a circularized half-light radius of 0$\farcs$91$\pm$0$\farcs$16, consistent with the curve-of-growth analysis. As the best-fit 2D model only has a peak signal-to-noise ratio (SNR) of 3, we run the following modelling to estimate the bias and the scatter of our measurement. Random elliptical Gaussian 2D models are first convolved with the synthesized beam and then injected at random positions on the residual averaged CO map with the best-fit model of the detected signals subtracted. {\sc imfit} is performed on each injected model and the output results are recorded. In total we inject 36000 models with the peak SNR, major axis, minor axis, and positional angle all randomized, in which the peak SNR has a range of 1 to 10 and the half-light radius in major and minor axis is allowed within 2$''$. We collect input parameters that correspond to an output matching to the CO measurements in peak SNR and circularized radius, and we compare the input and output circularized radius by computing the fractional difference defined as (output-input)/input. We find a 3\% upward bias in median (0.028$\pm$0.007) and a 20\% scatter. The scatter is consistent with the measurement, however the size bias needs to be corrected. We therefore obtain a bias-corrected $^{12}$CO($J$=3-2) half-light radius of 0$\farcs$88$\pm$0$\farcs$16. The deconvolved circularized half-light radius is therefore $0\farcs84\pm0\farcs16$. 

For the {\it uv}-based analyses we extract the averaged visibility over the FWHM channels as a function of the {\it uv} distances and then perform $\chi^2$ fitting assuming a Gaussian profile. We obtain a half-light radius of $0\farcs76\pm0\farcs10$, in good agreement with the result based on the image-plane analyses. However the {\it uv}-based measurement is better constrained with lower errors. We therefore adopt the {\it uv}-based measurement for the $^{12}$CO size. At the measured CO redshift, the $^{12}$CO half-light radius would therefore be $r_{1/2,{\rm CO}}=6.5\pm0.9$\,kpc. More details on the uv-fitting will be presented in Calistro Rivera et al.\ in preparation.

\subsection{SINFONI spectra}\label{subsec:sinfoni}
The H$\alpha$ line is strongly detected in our SINFONI data with a SNR$\sim$10 at the peak, allowing us to derive 2D intensity, velocity, and dispersion maps. In the following we therefore analyse the spectra in both integrated and 2D. Weaker [N\,{\sc ii}] and [S\,{\sc ii}] lines are also detected, although only in the central regions of the source. We also search for [O\,{\sc i}]6300, and [O\,{\sc iii}]/H$\beta$ at $\sim$5000\,\AA, but no significant detections are found.

In both 1D and 2D cases, we perform minimising-$\chi^2$ fit to the spectra over a wavelength range of 1.9--2.2\,$\mu$m, where the continuum is well described with a power-law slope and covers all the detected lines. The spectra are fit with four Gaussian models, in which all include a linear continuum component with the slope and normalization allowed free. We then fit different combination of lines; H$\alpha$, H$\alpha +$[N\,{\sc ii}], H$\alpha +$[S\,{\sc ii}], and H$\alpha +$[N\,{\sc ii}]$+$[S\,{\sc ii}]. In all cases we assume that the [N\,{\sc ii}] and [S\,{\sc ii}] lines have the same line width and redshift as those of H$\alpha$, and the flux ratio of the [N\,{\sc ii}]6583/[N\,{\sc ii}]6548\footnote{The values are air wavelengths and we use these for the ease of comparisons with the literature. Since SINFONI is situated in a cryo-vacuum chamber in the fitting process we adopt the vacuum wavelengths for [O\,{\sc i}] at 6302.1\,\AA, H$\alpha$ at 6564.7\,\AA, N{\sc ii} doublets at 6550.0\,\AA \, and 6585.4\,\AA, and [S\,{\sc ii}] doublets at 6718.4\,\AA\, and 6732.8\,\AA. These values are derived based on the conversion equation from air to vacuum wavelengths described in Equation 65 of \citet{Greisen:2006aa}.} doublet is fixed to a theoretical value of 3 based on the transition probabilities provided in \citet{1989agna.book.....O}. The flux ratio of the [S\,{\sc ii}]6731/[S\,{\sc ii}]6716 doublet is sensitive to the magnetic field strength hence it is not fixed. The fits are weighted against the sky spectrum provided by \citet{Rousselot:2000aa} and when calculating $\chi^2$ the wavelength ranges corresponding to the skylines are masked. The velocity dispersion is corrected in quadrature for instrumental broadening. The errors are derived using Monte Carlo simulations similar to those used for measuring the errors of the CO emission. Note that by adding an extra broad Gaussian component we have also searched for broad lines with a FWHM over 1000\,km\,s$^{-1}$, typical for SMGs hosting AGN and suggesting strong outflows (e.g., \citealt{Harrison:2012aa}), however we do not find evidence of such a broad component in ALESS67.1.

The model selection is determined based on the Akaike information criterion. Specifically we use the version that is corrected for a finite sample size (AICc; \citealt{Hurvich:1989aa}), which is defined as ${\rm AICc}=\chi^2+2k+2k(k+1)/(n-k-1)$, where $\chi^2$ is the $\chi^2$ from the fit, $k$ and $n$ denote the number of parameters and the number of data, respectively. Normally fits with more model parameters have lower $\chi^2$, and therefore a situation of over-fitting may not be reflected if one simply selects the model that produces the lowest $\chi^2$. The AICc offers a quantitative way to compare related models on the goodness of fit by penalising the number of parameters in the model, and the model that has the lowest AICc is selected as the adopted model in most cases. However, as shown below [N\,{\sc ii}]6583 happens to sit on one of the bright skylines, and we find that during the curve-of-growth analyses on the integrated spectra the skyline contamination becomes significant at larger radii. Consequently we restrict the fit to H$\alpha +$[N\,{\sc ii}]$+$[S\,{\sc ii}] only. 

Finally, a fit with line components is considered significant if the fit, compared to a simple continuum-only model, has a lower AICc and provides a $\chi^2$ improvement of $\Delta\chi^2>25$, equivalent to a SNR $>5$\,$\sigma$ assuming Gaussian noise and that the noise is not correlated among wavelength channels. For the models that include [S\,{\sc ii}], given it is a separated line without skyline contamination, we require a further $\chi^2$ improvement of $\Delta\chi^2>9$ (3\,$\sigma$) compared to the models without [S\,{\sc ii}]. 

\subsubsection{Integrated spectrum}\label{sec:sinfoni_tot}

\begin{figure}
	\begin{center}
		\leavevmode
		\includegraphics[scale=0.47]{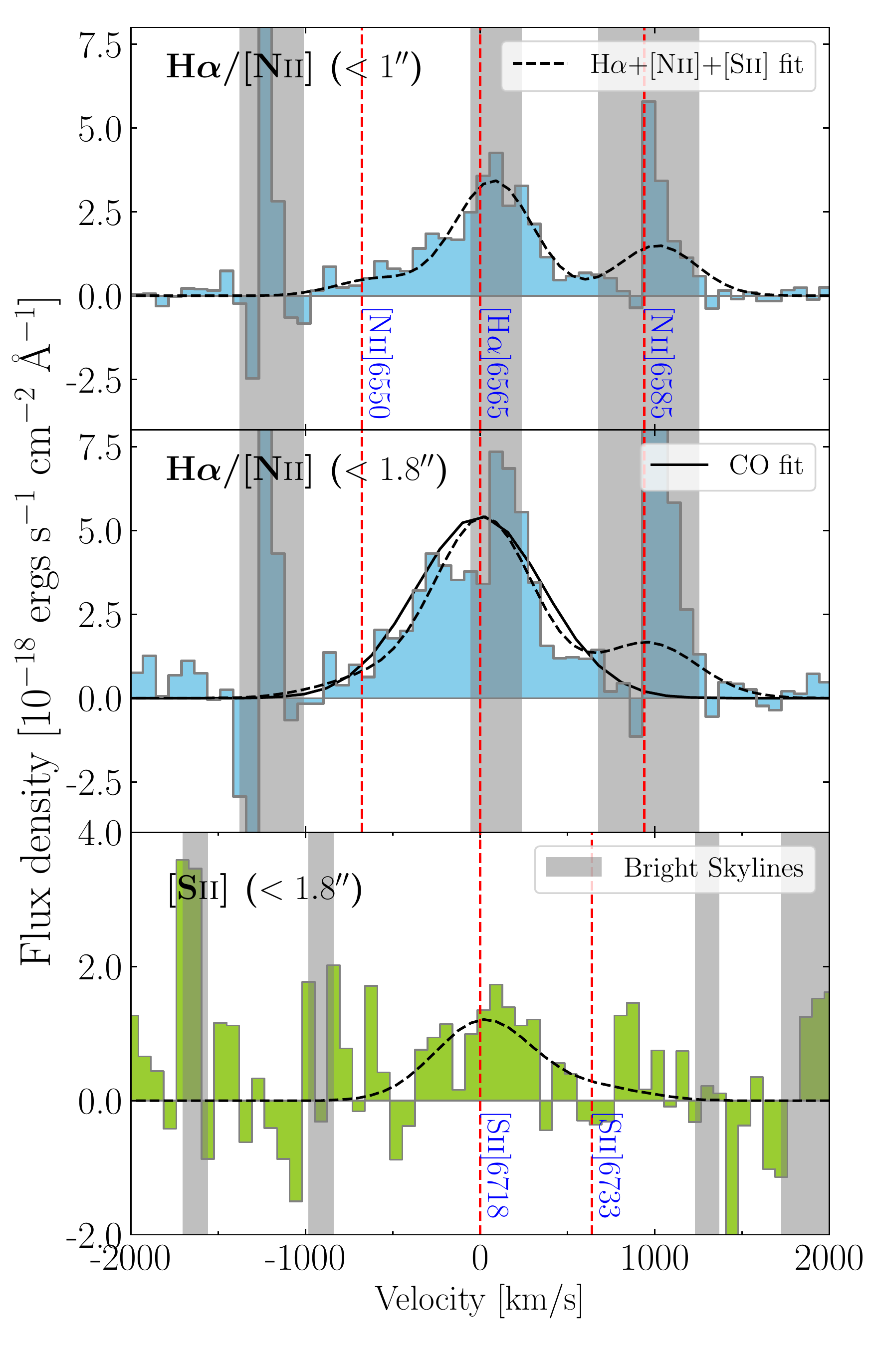}
		\caption{The integrated line profiles for CO, H$\alpha$/[N\,{\sc ii}], and [S\,{\sc ii}]. The top panel shows the integrated spectrum with an aperture radius of $1''$, in order to clearly show the detection of [N\,{\sc ii}], and the remaining two panels show the spectra with an aperture radius of $1\farcs$8, adopted based on the curve-of-growth analyses in which all the line fluxes are converged. The systemic velocity is referenced at CO and optical line redshifts, respectively. The grey vertical bands mark the positions of the bright sky lines, and the relative line positions are also marked according to their wavelengths. The best-fit CO Gaussian profile is shown as solid black curves in the top panel, and the best-fit line profiles for the optical lines are shown as dashed curves. The line profiles are consistent among the molecular and atomic emission lines, suggesting that in the integrated sense the dynamics that are measured by these tracers agree with each other.
		}
		\label{halpha}
	\end{center}
\end{figure}

To determine the radius over which the total flux is measured, we again employ a curve-of-growth approach that is similar to the one used for $^{12}$CO (see \autoref{sec:co}). We adopt the peak position of the 870\,$\mu$m continuum for the centroid, which produces a converged result and lies close to the geometrical center of the 2D intensity distributions shown in the next section. We again move the centroid around within the resolution area and fold the variations into the uncertainties of the measurements. The flux is derived based on the fitting procedure outlined in \autoref{subsec:sinfoni}, except at the radii larger than 1$\farcs$5, in which we find that the skyline contamination affects significantly on the fit and by examining the 2D maps we conclude that the [N\,{\sc ii}] lines are boosted (since there is no strong emission detected in the 2D map beyond this radius). We therefore fix the peak of the [N\,{\sc ii}] lines to be the one measured at 1$\farcs$5 but still allow dispersion and redshift to float. 

Based on the curve-of-growth approach, the integrated spectra are measured using a circular aperture with a radius of 1$\farcs$8, at which all three line fluxes are converged. From this, we measure a redshift of $z=2.1228\pm0.0006$, slightly higher but still consistent with the CO redshift within 3\,$\sigma$. 
We also measure a total H$\alpha$ flux of 2.6$\pm$0.4$\times$10$^{-16}$\,ergs s$^{-1}$ cm$^{-2}$, a [N\,{\sc ii}] flux of 1.1$\pm$0.4$\times$10$^{-16}$\,ergs s$^{-1}$ cm$^{-2}$, and a [S\,{\sc ii}] flux of 6.7$\pm$2.9$\times$10$^{-17}$\,ergs s$^{-1}$ cm$^{-2}$, with a spectral FWHM of 670$\pm$100 km s$^{-1}$. At the measured H$\alpha +$[N\,{\sc ii}]$+$[S\,{\sc ii}] redshift we compute a H$\alpha$ luminosity of $L_{\rm H\alpha}=9.2\pm1.5\times10^{42}$\,ergs s$^{-1}$,  a [N\,{\sc ii}] luminosity of $L_{\rm [N\,{\sc II}]}=3.8\pm1.4\times10^{42}$\,ergs s$^{-1}$, and a [S\,{\sc ii}] luminosity of $L_{\rm [S\,{\sc II}]}=2.4\pm1.1\times10^{42}$\,ergs s$^{-1}$.

The continuum-subtracted integrated spectra along with the best-fit Gaussian model are shown in \autoref{halpha}, in which we also show the best-fit $^{12}$CO($J$=3-2) profile for comparison. Because the spatial extend of [N\,{\sc ii}] is much smaller than that of H$\alpha$, the integrated spectrum with an aperture radius of 1$\farcs$8 includes extra unnecessary noises and [N\,{\sc ii}] may appear undetected. To demonstrate that [N\,{\sc ii}] lines are indeed detected in \autoref{halpha} we also plot the H$\alpha$/[N\,{\sc ii}] portion of the spectrum with a smaller aperture radius. The line profiles are consistent among the molecular and atomic emission lines, suggesting that in the integrated sense the dynamics that are measured by these tracers agree with each other. We compare in more detail the spatially-resolved dynamics between the two tracers in the discussion section. 

\begin{figure*}
	\begin{center}
		\leavevmode
		\includegraphics[scale=0.55]{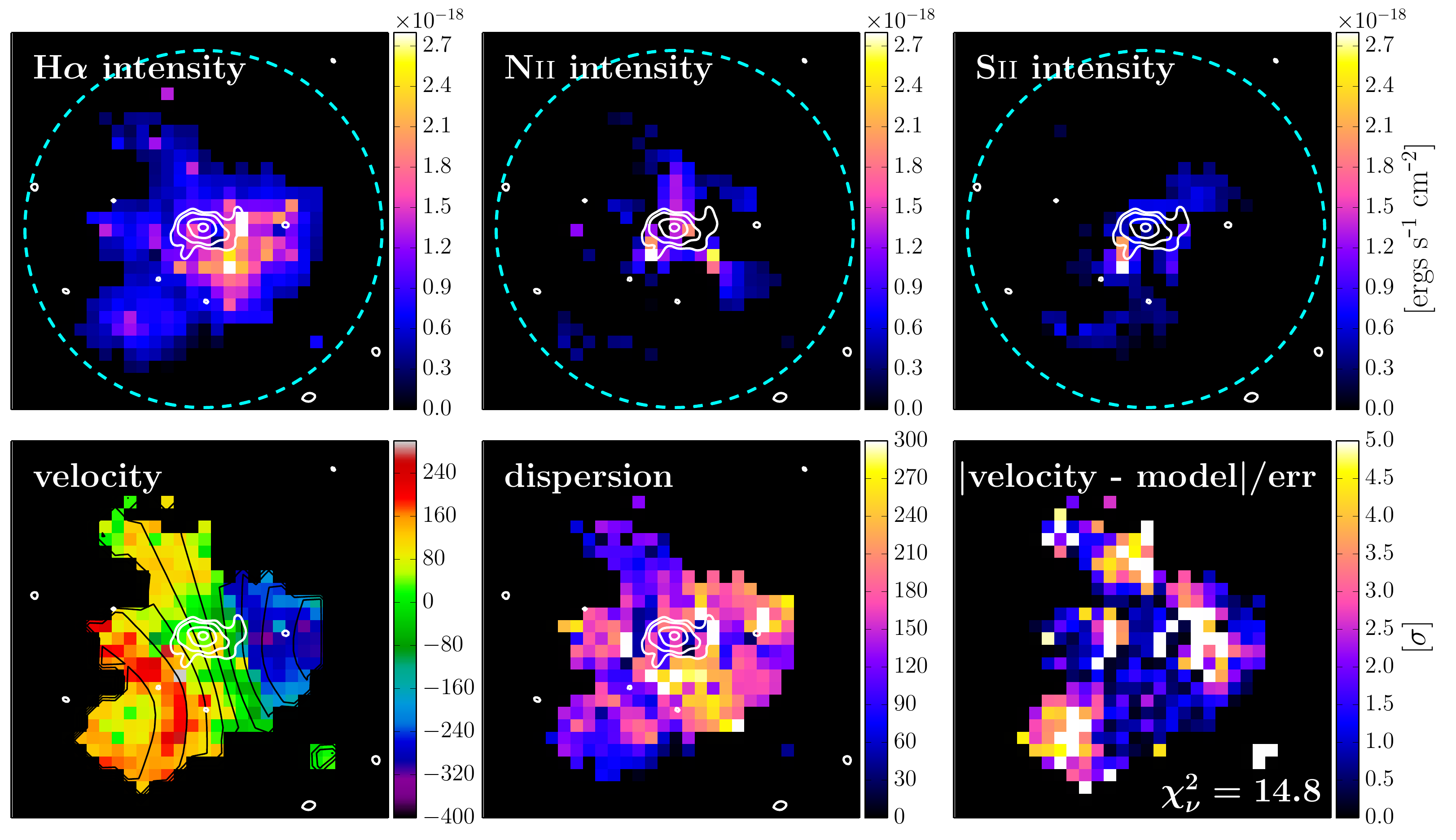}
		\caption{The 2D intensity maps for H$\alpha$, [N\,{\sc ii}], and [S\,{\sc ii}] shown in the top three panels with the same intensity scale. The cyan dashed circles represent the $1\farcs8$-radius circular aperture used to measure the total fluxes. All panels are overlaid with dust continuum in white contours, at levels of [3,5,10,15]$\times\sigma$. Strikingly, the peak of the H$\alpha$ and dust emission are not co-located, with H$\alpha$ also much more extended than the dust by a factor of $\sim$3. On the other hand the peaks of the [N\,{\sc ii}] and [S\,{\sc ii}] emission appears to match to that of dust emission, although both  are slightly more extended than dust. The bottom panels show the velocity field (bottom-left), velocity dispersion (bottom-middle), and the residual (bottom-right) in signal-to-noise between the measured velocity and a rotating disk model (\autoref{sec:2d}) with a reduced $\chi^2$ indicating a poor fit. The best-fit rotating disk model is plotted in the velocity map as black curves. We find that the velocity field of the optical emission lines is not consistent with orderly rotating disk.
		}
		\label{2Dmaps}
	\end{center}
\end{figure*}

The curve-of-growth analysis suggests a half-light radius of $0\farcs8\pm0\farcs1$ for H$\alpha$, which is slightly larger but consistent with $0\farcs63\pm0\farcs10$ derived from a best-fit 2D Gaussian profile on the intensity. We adopt the result from the curve-of-growth analysis since the projected H$\alpha$ emitting area is non-Gaussian with clear extended structures (\autoref{2Dmaps}) so a single Gaussian model is likely to underestimate the true size. Given the angular resolution of 0$\farcs$2 of the SINFONI observations, the deconvolved size of H$\alpha$ is $0\farcs77\pm0\farcs10$, corresponding to a H$\alpha$ half-light radius of $r_{1/2,{\rm H\alpha}}=6.6\pm0.9$\,kpc at the H$\alpha$ redshift. We perform the same curve-of-growth exercise for [N\,{\sc ii}] and [S\,{\sc ii}], finding $r_{1/2,{\rm NII}}=5.1\pm1.7$\,kpc and $r_{1/2,{\rm  SII}}=5.1\pm2.1$\,kpc. 

\subsubsection{Two-dimensional kinematics} \label{sec:2d}

To produce 2D intensity, velocity, and dispersion maps we run our line-fitting procedures described in \autoref{subsec:sinfoni} on each spaxel. However not every spaxel has significant line emission so we adopt an adaptive binning approach that is typically used for high-redshift IFU data (e.g., \citealt{Swinbank:2006aa}). We start with one spaxel, and if the fit is not significant then we average over $3\times3$ spaxels, and if that is still not significant then we increase the binning to $5\times5$ spaxels. In regions where this adaptive binning process still fails after $5\times5$ binning to give an adequate SNR, we leave the spaxel without a fit. The caveat of this approach is that the signals are weighted toward the higher SNR pixels.

The results are plotted in \autoref{2Dmaps}, showing the 2D intensity maps for H$\alpha$, [N\,{\sc ii}], and [S\,{\sc ii}]. The velocity and the velocity dispersion map are also shown.

The first and most striking feature is how most of the H$\alpha$ and dust emission are not co-located, with H$\alpha$ much more extended than the dust by a factor of $\sim$3. The sky separation between the peaks of the H$\alpha$ and 870\,$\mu$m continuum is 0$\farcs$4, more than 3\,$\sigma$ given $\sim$0$\farcs$2 resolution in FWHM for both the ALMA and SINFONI observations. Although the systematic uncertainty in SINFONI astrometry could contribute to a further offset of $\sim0\farcs2-0\farcs3$, later we show that the cold dust emission coincides with the regions with the reddest colors revealed by the WFC3 imaging, and most of the  H$\alpha$ emission matches the location of the brightest continuum in rest-frame optical WFC3 maps. Therefore we conclude that the apparently disjoint nature between the cold dust as traced by the 870\,$\mu$m continuum and the H$\alpha$ emission is genuine. However, on the other hand, we find that the sky locations of [N\,{\sc ii}] and [S\,{\sc ii}] peak at the position of the dust emission, although both [N\,{\sc ii}] and [S\,{\sc ii}] are slightly more extended than dust. The enhanced [N\,{\sc ii}]-to-H$\alpha$ ratio in the central regions could indicate higher gas-phase metallicity. However the detected X-ray emission toward ALESS67.1 could also suggest that the elevated ratio is caused by the harder radiation field and/or shocks from AGN. Deeper observations with detections including [O\,{\sc iii}] and H$\beta$ should help distinguish between these alternatives.

Lastly, the 2D velocity map shown in \autoref{2Dmaps} displays a velocity gradient in H$\alpha$ kinematics, with a peak-to-peak velocity difference of 750$\pm$220\,km s$^{-1}$. Given the integrated velocity dispersion of 280$\pm$40\,km s$^{-1}$ (\autoref{sec:sinfoni_tot}), ALESS67.1 would be considered rotation-dominated based on the criterion of $(v_{max}-v_{min})/2\sigma_{int} = 0.4$, which has been used in some work to roughly differentiate orderly rotating disk and merger (e.g., \citealt{Forster-Schreiber:2009aa}). However, the availability of the 2D velocity map allows us to conduct detail kinematic modelling to more reliably differentiate rotating systems from mergers. 

We start by modelling the velocity field assuming a rotating disc. We adopt the simplest function for the rotational curve, the arctan function \citep{Courteau:1997aa}, with the one dimensional (1D) form of 

\begin{equation}\label{RC}
v(r) = v_0 + \frac{2}{\pi}v_{asym} \textrm{arctan}\Big(\frac{r-r_0}{r_t}\Big)
\end{equation}
where $v_0$ is the systemic velocity, which in our case is zero as the velocity field is referenced at the systematic redshift, $v_{asym}$ is the asymptotic velocity,  $r_0$ is the central position and $r_t$ is the transition radius between the rising and flat part of the rotational curve. The arctan function has been found to have the flexibility to reasonably describe $z\gtrsim1$ rotating galaxies (e.g., \citealt{Miller:2011aa, Swinbank:2012aa}). Based on Appendix A of \citet{Begeman:1989aa} the 1D rotational curve is projected to 2D via

\begin{equation*}
\begin{split}
v_p(x, y) = v(x,y)\textrm{sin}(i) 
\frac{-(x-x_0)\textrm{sin}\phi+(y-y_0)\textrm{cos}\phi}{\sqrt{(x-x_0)^2+(y-y_0)^2}}
\end{split}
\end{equation*}
where $i$ is the inclination angle, the angle between the normal to the plane of the galaxy and the line of sight (i.e. 0$^{\circ}$ if face-on and 90$^{\circ}$ if edge-on), $x_0$ and $y_0$ is the central sky position, and $\phi$ is the positional angle (P.A.) of the major axis, defined as the angle taken in anti-clockwise direction between the north direction in the sky and the major axis of the receding half of the galaxy. 

We fit the 2D model to the measured data based on maximum likelihood, in particular we run {\sc emcee}, a Markov chain Monte Carlo (MCMC) ensemble sampler \citep{Foreman-Mackey:2013aa}, to explore the parameter space and derive uncertainties. We limit the parameters to the range that is allowed by the data, in particular, the center position ($x_0$/$y_0$) and the turnover radius ($r_t$) must be within the SINFONI field-of-view and the P.A. lies between 0$^{\circ}$ and 180$^{\circ}$. Because $v_{asym}$ and $i$ are essentially degenerate for our data quality we treat $v_{asym}{\rm sin}(i)$ as a single parameter.

The best-fit model has a $v_{asym}{\rm sin}(i) = 290\pm70$\,km s$^{-1}$ and a P.A. of $110\pm20$ degrees, and is plotted over the velocity field in \autoref{2Dmaps}, in which the difference between the model and the data is also shown. The reduced $\chi^2_\nu=14.8$, indicating a relatively poor fit to the data and suggesting ALESS67.1 is not a pure rotating disk.

While modelling the 2D velocity field offers clues to whether the system is well-described by a simple rotating disk, quantifying the asymmetry of both velocity and velocity dispersion provides a more complete and well-defined view of the kinematics of the system. One well-tested way to measure the asymmetry of the kinematics is to use kinemetry, originally presented by \citet{Krajnovic:2006aa} and designed to study local high SNR stellar kinematic data such as those from the SAURON project \citep{Bacon:2001aa}. It has been further developed into an effective tool to separate disks from mergers (e.g., \citealt{Shapiro:2008aa}), although at $z>1$ the effectiveness may depend on the interaction stage of the merger (e.g., \citealt{Hung:2015aa}).

\begin{figure*}
	\begin{center}
		\leavevmode
		\includegraphics[scale=0.58]{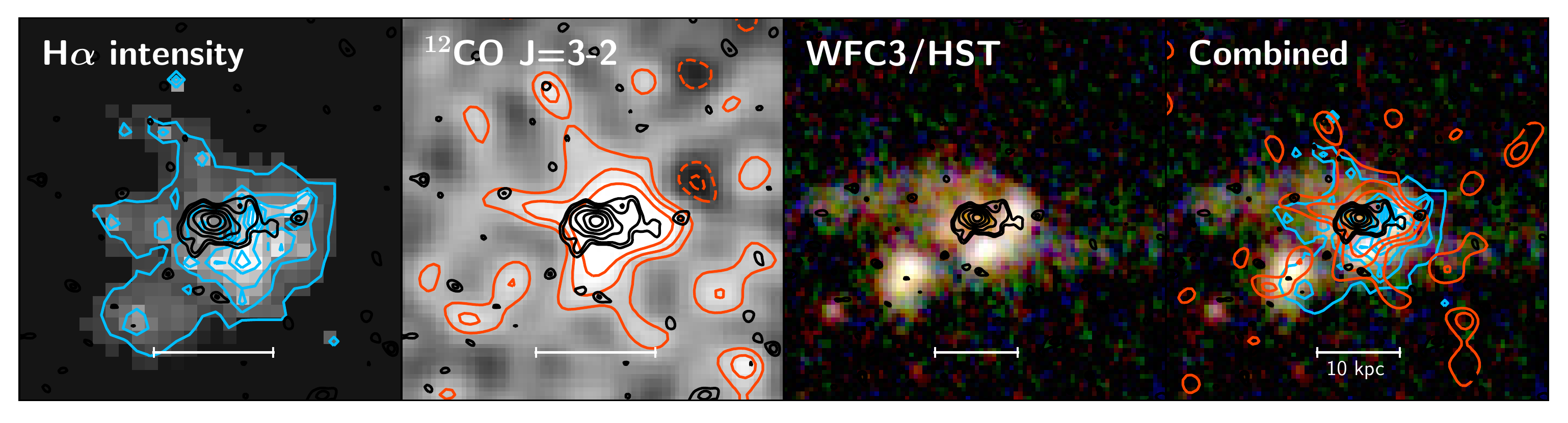}
		\caption{Thumbnails of ALESS67.1 in H$\alpha$, CO, NIR {\it HST} r-g-b (F160W-F125W-F105W) imaging, and the combination of all, with black contours showing the high-resolution 870\,$\mu$m continuum at the same levels as \autoref{cocont}. The white horizontal bars mark the physical scale of $\sim$10\,kpc. Note the left two panels are slightly zoomed in to better show the detailed structures in the central regions. Note the different structures revealed by different tracer, cautioning the assumption of single size or morphology when assessing the galaxy properties.
		}
		\label{all}
	\end{center}
\end{figure*}

The basic goal of kinemetry analysis is to first decompose the 2D kinematic moment (e.g., velocity and velocity dispersion) maps into a series of concentric ellipses with increasing major axis length, which are defined by the systemic center and positional angle \citep{Krajnovic:2006aa}. For each concentric elliptical ring the kinematic moments as a function of the azimuthal angle is then extracted and further decomposed into the Fourier series, which can be described as

\begin{equation*}
K(r, \psi) = A_0(r) + \sum_{n=1}^{N}A_n(r)\textrm{sin}(n\psi)+B_n(r)\textrm{cos}(n\psi)
\end{equation*}
where $r$ is the length of the semi-major axis in each elliptical ring, $\psi$ is the azimuthal angel, and $A_n$ and $B_n$ are $n$th-order coefficients. This equation can be shortened as 

\begin{equation*}
K(r,\psi) = A_0(r) + \sum_{n=1}^{N}k_n(r)\textrm{cos}\{n\big[\psi-\phi_n(r)\big]\}
\end{equation*}
where $k_n = \sqrt{A_n^2+B_n^2}$ and $\phi_n = \textrm{arctan}\big(A_n/B_n\big)$.

In the case of an ideal rotating disk, the kinemetry of the velocity and velocity dispersion field would be dominated by $B_1$ and $A_0$ coefficients, respectively. Any perturbation from an ideal disk would manifest itself in the higher-order kinemetry coefficients. Therefore the ratios between the high-order and the dominant coefficients in the ideal disk case can be used to quantify the kinematic asymmetry. By using a sample of local galaxies as the training sample \citet{Shapiro:2008aa} proposed a criteria of $K_{asym} = (v_{asym}^2+\sigma_{asym}^2)^{1/2}=0.5$ to separate rotating disk ($K_{asym} < 0.5$) and merger ($K_{asym} > 0.5$), where 
\begin{equation*}
\begin{split}
v_{asym} = \Big\langle\frac{k_{2,v}+k_{3,v}+k_{4,v}+k_{5,v}}{4B_{1,v}}\Big\rangle_r, \\
\sigma_{asym} = \Big\langle\frac{k_{1,\sigma}+k_{2,\sigma}+k_{3,\sigma}+k_{4,\sigma}+k_{5,\sigma}}{5B_{1,\sigma}}\Big\rangle_r.
\end{split}
\end{equation*}

We run the kinemetry code provided by \citet{Krajnovic:2006aa} on the velocity and velocity dispersion maps shown in \autoref{2Dmaps}. As pointed out by \citet{Krajnovic:2006aa} the dominant uncertainty of the kinemetry analyses is the choice of the center position. We therefore adopt the best-fit position based on the previous rotational curve modelling but perturb it within the error obtained from the MCMC analyses. We find a median $v_{asym} = 0.15\pm0.01$ and a median $\sigma_{asym} = 0.48\pm0.13$, which leads to $K_{asym} = 0.64\pm0.15$, suggesting that judging from the optical line kinematics ALESS67.1 is a borderline merger (still consistent with a rotating disk given the error) based on the criteria proposed by \citet{Shapiro:2008aa}. 

\section{Discussion}\label{sec:discussion}

We have presented detailed analyses of ALMA and SINFONI observations of the $z=2.12$ SMG ALESS67.1, which provide information on cold dust continuum, molecular gas, and atomic emission lines predominantly coming from H{\sc ii} regions, all with sub-arcsecond resolution. All of these tracers are resolved in our data and the structures of each tracer are revealed. We compare these structures by overlaying each tracer on top of one another in \autoref{all}, where we also include the WFC3 imaging from {\it HST}, which predominantly show the distribution of unobscured stars. The difference in size and spatial distribution among each tracer, the main finding of this paper, is clear. The availability of all these data means that we can also test some model predictions, as well as attempt to explain recent findings on $z\sim2$ galaxies regarding the dust attenuation and the deviation of the star-formation law. Lastly, the fact that these tracers are not co-spatial may have a profound impact on subjects such as galaxy SED modelling, which typically assumes that all components are co-located. In the following we discuss these implications.

\subsection{Kinematics -- CO and H$\alpha$}\label{sec:kin}
Measuring the kinematics of a galaxy provides an insightful view on its formation and evolution history. For SMGs in particular, the kinematic measurements have mostly been used to assess the mechanisms that drive the enhancement of star formation (e.g., \citealt{Bouche:2007aa}) and to investigate their evolutionary link to the local massive ellipticals (e.g., \citealt{Swinbank:2006aa}).

\begin{figure}
	\begin{center}
		\leavevmode
		\includegraphics[scale=0.65]{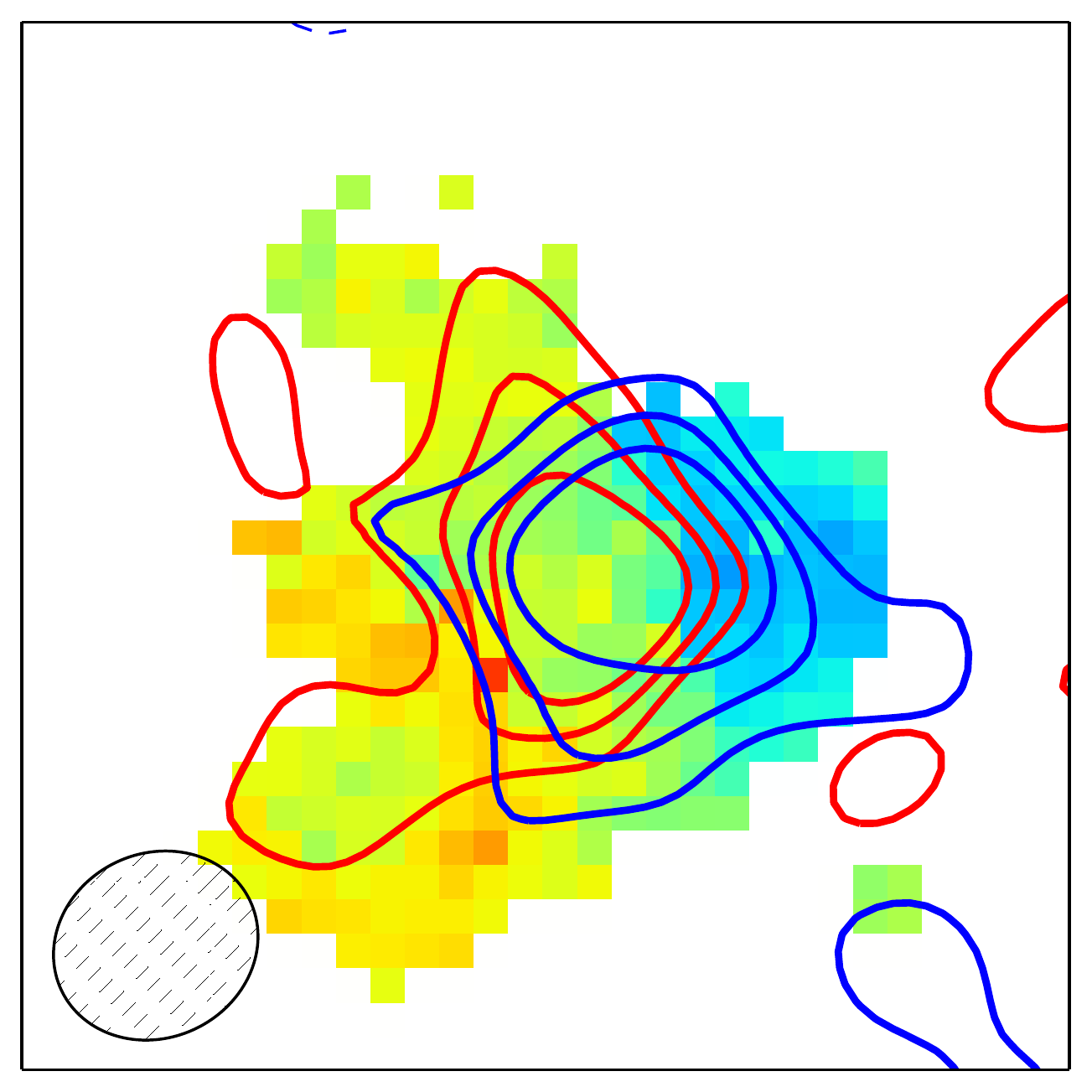}
		\caption{A comparison between velocity of the optical lines derived from SINFONI with that of CO from ALMA overlaid as contours. The approaching side is shown in blue and receding side in red. The synthesized beams are plotted at the bottom-left corner, which has a FWHM of $\sim0\farcs7$ in major axis. The contours have levels at [-3,-2,2,3,4]$\times \sigma$, in which the positive and negative values are plotted as solid and dashed curves, respectively. The mean velocities of the two CO channels are $\pm$215 km s$^{-1}$, comparable to the optical-line dynamics. We find that the kinematics of both HII regions (optical strong lines) and molecular gas (CO) are in broad agreement. 
		}
		\label{pv}
	\end{center}
\end{figure}

However, their high dust obscuration and low surface density means that it is difficult to obtain $\sim$\,kpc resolution optical emission line observations, which require AO for $z\sim2$ SMGs and are essential to complement the kinematics measured using molecular gas made with ALMA. Consequently there are still only a handful of SMGs that have AO-aided strong optical line data (e.g., \citealt{Menendez-Delmestre:2013aa, Olivares:2016aa}). Together with the technical difficulty of obtaining sub-arcsecond resolution data for (sub)millimeter molecular lines, in particular low-$J$ CO, most studies have relied on single tracer to assess and compare the kinematics, assuming that different tracers behave similarly. With sub-arcsecond resolution data in hand for both $^{12}$CO(J=3-2) and H$\alpha$ we are in position to compare these tracers in kinematics.

Our $^{12}$CO(J=3-2) detection does not have high enough SNR to produce 2D kinematic maps like H$\alpha$, we therefore taper the CO map, divide the line in half, and show the averaged maps in both the receding and approaching side. We overlay the results on the 2D velocity map obtained from SINFONI in \autoref{pv}.

As revealed in \autoref{pv}, in the outskirts the extended CO emission appears to follow the bulk rotating motion of the optical emission lines, in particular in the south-east filament. This filament links to a second stellar component revealed in the {\it HST} imaging (\autoref{all}) and could suggest that this second stellar component along with the connecting gas stream was falling toward the dynamical center close to the 870\,$\mu$m continuum. The peak of the blue-shifted part of the CO line agrees with that of the red-shifted part, however higher SNR CO observations are needed to reveal the detailed kinematics in the central regions. In general we find that the kinematics of both HII regions as traced by the optical emission lines and molecular gas as traced by $^{12}$CO(J=3-2) are in broad agreement.

\subsection{$\alpha_{\rm CO}$}\label{sec:alphaCO}
The interstellar medium (ISM) of the dense environments in which rapid star formation occurs are dominated by molecular hydrogen, H$_2$.  As a result H$_2$ plays a central role in the formation and evolution of galaxies \citep{Kennicutt:2012aa}. While cold H$_2$ is not directly observable in emission, $^{12}$CO has been widely used to trace total molecular gas, with a standard conversion between $^{12}$CO luminosity and the total molecular gas mass, M$_{\rm mol} = \alpha_{\rm CO} L^\prime_{\rm CO(1-0)}$, in which $\alpha_{\rm CO}$ is the CO-to-H$_2$ conversion factor. For extragalactic sources, the conversion factor depends on the properties of the galactic environments such as gas density and metallicity \citep{Bolatto:2013aa}, resulting in a factor of $\sim$6 variation between the local mergers and the local spiral galaxies \citep{Downes:1998aa}.

Previous studies have attempted to quantify $\alpha_{\rm CO}$ for SMGs, however subject to a lack of constraints on dust mass and/or the size and dynamics of the CO emission, the results were generally not conclusive (e.g., \citealt{Bothwell:2013lp}), though on average SMGs were found to have a low $\alpha_{\rm CO}$ \citep{Danielson:2011aa}. With all the necessary measurements in hand for a case of merger we can quantify and test the link between merger and $\alpha_{\rm CO}$.

To measure $\alpha_{\rm CO}$ we first adopt a dynamical method, in which the gas mass is derived by subtracting stellar and dark-matter mass from the dynamical mass (i.e., $M_{\rm dyn}(r\leq r_e) = 0.5\times(M_{\rm star}+M_{\rm gas})+M_{\rm dark}(r\leq r_e)$, where $r_e$ represents half-light radius and we assume the gas and stars have the same r$_e$ given this is what we find (\autoref{cog_all}). Within the half-light radius the dark matter contribution (M$_{\rm dark}$) is estimated to be 10--20\%  of the dynamical mass in $z\sim2$ star-forming galaxies (e.g., \citealt{Genzel:2017aa}) so we adopt 15\%. Since we find that ALESS67.1 is rotation-dominated based on the optical emission lines \autoref{sec:2d}, the dynamical masses can be estimated using the Newtonian dynamics assuming a point mass, $v_c(r_e) = \sqrt{GM_{\rm dyn}(<r_e)/r_e}$, in which $v_c$ is the circular velocity and $G$ is the gravitational constant. Based on the dynamical modelling presented in \autoref{sec:2d} $v_c$ can be described as \autoref{RC}. However in the highly turbulent environment ($v_{asym}$/$\sigma \lesssim 3$), the rotational velocity is significantly reduced due to the turbulent pressure effects and it needs to be corrected, with a form of $v_c^2(r_e) = v^2(r_e) + 3.36\times\sigma^2$ (Equation 11 in \citet{Burkert:2010aa} assuming exponential surface density distribution). This correction assumes constant velocity dispersion ($\sigma$) over the spatial extend, which is supported by our data (\autoref{2Dmaps}). We derive a median $\sigma$ of $154\pm5$km s$^{-1}$ with a bootstrapped error. By adopting an averaged inclination $\langle{\rm sin}^2(i)\rangle=2/3$ (Tacconi et al. 2008) and a half-light radius of $r_{1/2,{\rm H\alpha}}=6.6\pm0.9$\,kpc we derive a $v_c(r_e) = 380\pm40$ km s$^{-1}$ so a dynamical mass of $M_{\rm dyn}(r\leq r_e)=(2.2\pm0.6)\times10^{11}$M$_\odot$. The stellar mass is estimated to be $M_{\rm star}=2\times10^{11}$M$_\odot$ \citep{Simpson:2014aa, da-Cunha:2015aa}. Hence we estimate a gas mass $M_{\rm gas} = (1.8\pm1.0)\times10^{11}$M$_\odot$. Given $L^\prime_{\rm CO(1-0)}=(9.9\pm1.8)\times10^{10}$\,K\,km\,s$^{-1}$\,pc$^2$ \citep{Huynh:2017aa}, we derive an $\alpha_{\rm CO}=1.8\pm1.1$.

We can also check our $\alpha_{\rm CO}$ estimate by using a gas-to-dust ratio method. The idea is that by estimating the gas-to-dust mass ratio, $\delta_{GDR}$, in a galaxy with measured molecular gas and dust masses, the conversion factor $\alpha_{\rm CO}$ can be derived as $\alpha_{\rm CO} = \delta_{GDR}(\mu_0)M_{\rm dust}/L'_{\rm CO(1-0)}$ assuming that molecular gas dominates the gas masses, and $\delta_{\rm GDR}$ can be related to gas phase metallicity ($\mu_0$; \citealt{Leroy:2011aa}). Assuming negligible AGN contribution to the [N\,{\sc ii}]/H$\alpha$ ratio, we derive a metallicity of 12+log(O/H) = 8.8$\pm$0.1 in the central region, and 8.6$\pm$0.1 over the entire galaxy, by adopting the $N2\equiv$\,log$_{10}$([N{\sc ii}]$\lambda$6583/{\rm H$\alpha$}) empirical calibration provided by Pettini \& Pagel (2004) (12 + log$_{10}$(O/H) = 8.90 + 0.57$\times N2$). By using a circular aperture on the CO(J=3-2) cube centered at the 870\,$\mu$m continuum peak with a deconvolved radius matching to the dusty region ($\leq 0\farcs4$) we estimate $L^\prime_{\rm CO(1-0)} = (4.0\pm1.9)\times10^{10}$\,K\,km\,s$^{-1}$\,pc$^2$. Based on the metallicity measurement and the best-fit linear function provided by \citet{Leroy:2011aa} we compute a $\delta_{\rm GDR}$ of $78\pm6$ in the central region. By adopting the dust mass of ALESS67.1 derived by \citet{Swinbank:2014aa} we calculate a gas mass of $(5.5\pm0.6)\times 10^{10} M_\odot$, and therefore a $\alpha_{\rm CO,GDR}=1.4\pm0.7$ in the central dusty regions, in good agreement with the $\alpha_{\rm CO}$ estimated using the dynamical method. On the other hand, it is also possible that the apparent size difference between dust and CO is caused by the observational bias due to the different optical depth probed by CO and 870\,$\mu$m continuum. In such a case CO and dust are still well-mixed in the entire galaxy and the total CO luminosity needs to be adopted for the gas-to-dust ratio method. If we do so and adopt global integrated values for all relevant measurements, we would obtain a lower $\alpha_{\rm CO}$ of 0.5$\pm$0.3, consistent with the results derived using other methods. The results of the gas-to-dust ratio method should be treated as lower limits as the AGN contribution could lower the gas-phase metallicity and hence suggest higher gas masses.
	
Despite the uncertainty, our result is consistent with that of another strongly lensed SMG SMMJ2135-0102 (Cosmic Eyelash; \citealt{Danielson:2011aa,Danielson:2013aa,Thomson:2015aa}), and Arp 220 \citep{Scoville:1997aa}, a local merger that has been used to compare with $z\sim2$ SMGs, with both having $\alpha_{\rm CO}\sim1$. Confirming the results from other analyses, our results on $\alpha_{\rm CO}$ suggest a consistent scenario that ALESS67.1 is undergoing a merger. 

\subsection{Sizes and morphology}
\begin{figure}
	\begin{center}
		\leavevmode
		\includegraphics[scale=0.85]{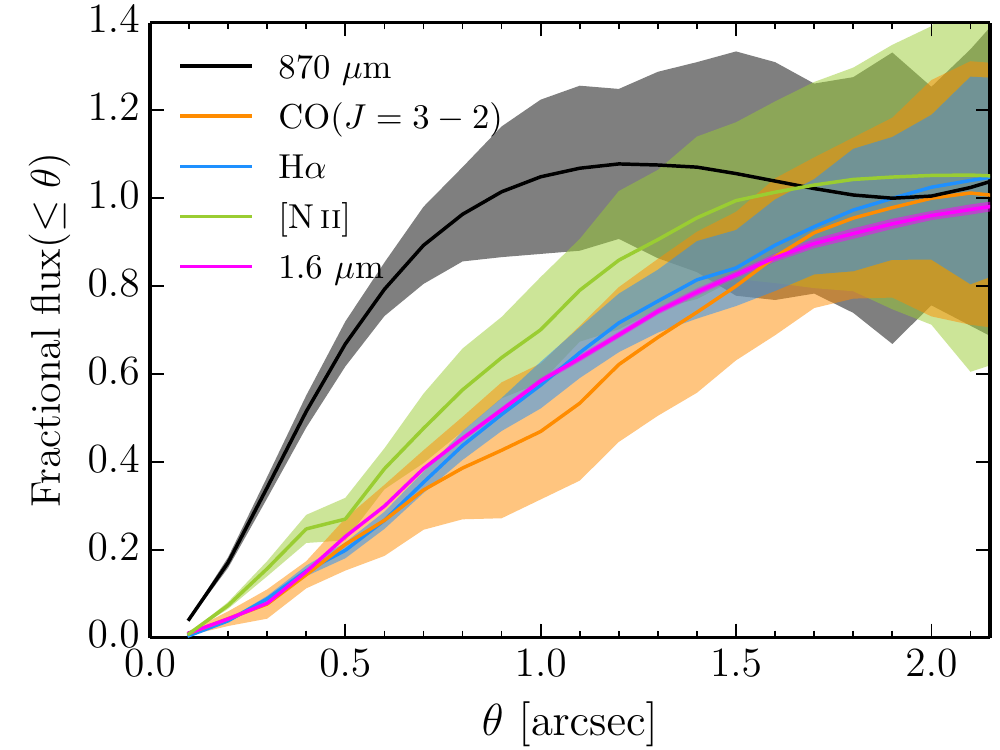}
		\caption{The curve-of-growth diagram with all the tracers plotted. All the maps are convolved with the CO beam to make the spatial resolution comparable. Note the size difference between 870\,$\mu$m and the rest of the tracers, especially CO. Our results caution against assumptions in any model or analyses that adopt a single geometry for all the tracers in SMGs.
		}
		\label{cog_all}
	\end{center}
\end{figure}
Recent SMG studies of kpc-scale dust distributions using ALMA (sub)millimeter observations have found, almost unequivocally, compact sizes with an average half-light radius of 1--2\,kpc (e.g., \citealt{Simpson:2015aa,Ikarashi:2015aa,Spilker:2016aa,Hodge:2016aa}).  Above an infrared luminosity of $\sim3\times10^{12}$\,$L_{\rm IR}$ (SFR$\sim$300\,M$_\odot$\,yr$^{-1}$), the dust sizes do not appear to depend on $L_{\rm IR}$ and redshift ($z\sim1-6$), and they are on average a factor of 2--3 smaller than the near-infrared continuum revealed by the {\it HST} (Hodge et al. 2016).

We have presented analyses and measured sizes and spatial distributions of CO, optical emission lines, 870\,$\mu$m and near-infrared continuum for ALESS67.1. We summarize part of the measurements in \autoref{cog_all} in a curve-of-growth style to emphasize the size difference among each tracer. All maps are convolved with the $^{12}$CO beam to make the spatial resolution comparable. 

\begin{figure*}
	\begin{center}
		\leavevmode
		\includegraphics[scale=1]{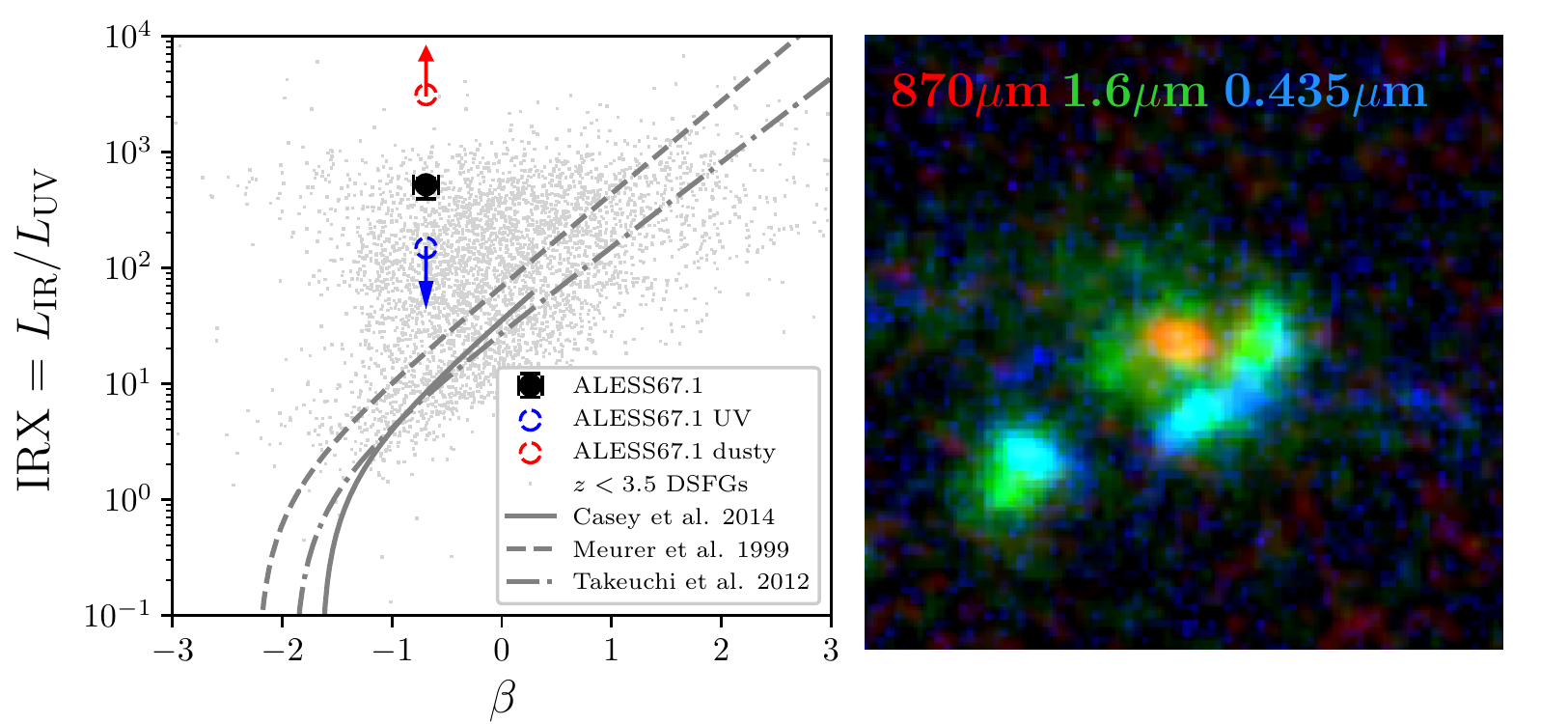}
		\caption{{\it Right:} The IRX-$\beta$ diagram with the black point representing ALESS67.1 and grey dots are $z<3.5$ DSFGs \citep{Casey:2014ab}. The various correlations found in the local SFGs are also plotted \citep{Meurer:1999aa,Takeuchi:2012aa,Casey:2014ab}. ALESS67.1 as a whole is significantly bluer (more IR luminous) compared to the local relationships, but locating at the locus of the $z<3.5$ DSFGs. We also plot the limits of both dusty regions revealed by the 870\,$\mu$m continuum and UV-emitting regions as red- and blue-dashed circle, respectively, by fixing to the same $\beta$ for the dusty regions. {\it Right}: False-color r-g-b(870\,$\mu$m-1.6$\mu$m-0.435$\mu$m) thumbnail, indicating a complete mismatch between cold dust and UV emitting regions, supporting the postulation that geometrical effect is part of the reason causing the deviation of DSFGs on the local IRX-$\beta$ correlations.
		}
		\label{IRXb}
	\end{center}
\end{figure*}

Besides confirming the size contrast between 870\,$\mu$m and 1.6\,$\mu$m continuum, we also find significantly larger sizes for all other tracers with respect to the 870\,$\mu$m dust continuum. Perhaps the most striking of all is the factor of 5.2$\pm$0.8 difference in half-light radius between FIR/submillimeter dust continuum and the  $^{12}$CO($J$=3-2) gas, which is normally found to agree within a factor of two in nearby star-forming galaxies (e.g., \citealt{Sandstrom:2013aa}), as well as local mergers such as Arp220 \citep{Scoville:1991aa}, NGC 6240 \citep{Iono:2007aa}, and NGC 3256 \citep{Sakamoto:2006aa}.

The measured $^{12}$CO($J$=3-2) size of ALESS67.1 is consistent with that of $^{12}$CO(2-1) from the $z=4.055$ SMG GN20 (Hodge et al. 2012), as well as that of $^{12}$CO(1-0) in a sample of four SMGs presented in \citet{Ivison:2011aa} and the $z=3.408$ SMG SMMJ13120+4242 \citep{Riechers:2011aa}. However, it is significantly larger than the $^{12}$CO sizes reported by \citet{Engel:2010p9470}, in which they claimed an average half-light radius of 2.5$\pm$1.3\,kpc for $^{12}$CO lines with upper $J\leq6$. The difference could be a reflection of true scatter in $^{12}$CO sizes, which is a factor of $\sim$8 in local mergers \citep{Ueda:2014aa}, or as claimed by \citet{Ivison:2011aa} that low-$J$ lines are more spatially extended than high-$J$ ones, or both.

On the other hand, in terms of spatial extend and surface brightness, we find that H$\alpha$ follows a similar distribution as the stellar components traced by the near-infrared emission. When comparing H$\alpha$/near-infrared continuum to $^{12}$CO we find similar sizes.  

When comparing to the measurements in the literature, the near-infrared size of ALESS67.1, which is 6.4$\pm$0.5\,kpc using the curve-of-growth method (\autoref{cog_all}), lies within the average size of the ALESS parent sample \citep{Chen:2015aa} and given the stellar mass estimates is consistent with the size census on mass-selected samples of star-forming galaxies at $z\sim2$ \citep{van-der-Wel:2014aa}. The near-infrared size of ALESS67.1 is however significantly larger than that of 500\,$\mu$m-selected DSFGs reported by \citet{Calanog:2014aa}, in which they attribute the difference to the selection bias caused by the strong gravitational lensing in their sample.

The H$\alpha$ size of ALESS67.1 lies at the higher end but still consistent with respect to other SMGs reported in \citet{Alaghband-Zadeh:2012aa} (average 3.7$\pm$0.8\,kpc) and main-sequence SFGs \citep{Forster-Schreiber:2009aa}, but significantly larger than the H$\alpha$ emitters (HAEs) at similar reshifts \citep{Molina:2017aa}. The difference between ALESS67.1 and HAEs, apart from small samples, could be caused by the fact that the HAE sample presented in Molina et al. is on average a factor of $>3$ less massive than ALESS67.1.

Lastly, the geometrical discrepancy among different tracers, which is also seen in other studies of high-$z$ galaxies (e.g., \citealt{Spilker:2015aa, Decarli:2016aa,Koprowski:2016ab,Ginolfi:2016aa}), may have some impact on theoretical modelling. As already discussed in 
\citet{Simpson:2017ab}, geometrical differences in SMGs between dust and UV-to-near-infrared emissions leads to drastically different estimations for dust extinction, with Simpson et al.\ deriving an $A_{\rm V}=540^{+80}_{-40}$ using a Hydrogen column density method based on direct measurements of dust column density, in contrast to just $A_{\rm V}\sim1-3$ based on SED modelling of the detectable optical emissions either using simple dust screen modelling (Simpson et al. 2014) or energy-balance approach (da Cunha et al. 2015). For SMGs, because of this geometrical discrepancy it may be more sensible to model optical-to-NIR SEDs and FIR/submillmeter/radio SEDs separately. 

In the next two sections we now discuss the impact of this geometrical discrepancy on the topic of IRX-$\beta$ relationship and the Schmidt-Kennicutt relationship.

\subsection{IRX-$\beta$}

The relationship between the ratio of infrared and UV luminosity at 1600\,\AA\, (IRX) and the UV spectral slope at 1600\,\AA\, ($\beta$) offers a potential route to estimate the total SFR when only the rest-frame UV observations are available. This IRX-$\beta$ relationship has therefore been widely used to estimate the total star-formation rate density for UV-selected populations at $z>3$ and up to the epoch of reionization (e.g., \citealt{Bouwens:2015aa}).

Given its fundamental implications for the measurements of SFR density at high redshifts, the IRX-$\beta$ relationship and the deviation of it has been extensively studied both in the local Universe and at high redshifts (e.g., \citealt{Meurer:1999aa, Kong:2004aa, Buat:2005aa, Howell:2010aa, Overzier:2011aa,Takeuchi:2012aa,Reddy:2012aa,To:2014aa, Casey:2014ab}). Among the many factors that affect the IRX-$\beta$ relationship such as star formation history and internal attenuation curve, geometrical effect has been proposed to explain the deviation seen in samples of the ultraluminous infrared galaxies, both in the local Universe \citep{Howell:2010aa} and at $z\sim2$ \citep{Casey:2014ab}, in a sense that the different or completely decoupled geometry between dust and UV could explain the increase of the IRX at a fix $\beta$. Recent theoretical models have also confirmed this hypothesis (e.g., \citealt{Narayanan:2017aa, Popping:2017aa}). With $\sim$\,kpc resolution data in rest-frame UV, optical and FIR we are set to examine this proposal.

We first compute the UV luminosity and the spectral slope at 1600\,\AA. ALESS67.1 is covered by the CANDELS imaging (\autoref{all}) as well as 3D-{\it HST} grism spectroscopy \citep{Brammer:2012aa}, and the multi-wavelength photometry and rest-frame UV flues at 1400, 1700, 2200, 2700, 2800\,\AA\, are provided in \citet{Skelton:2014aa}. We adopt the values in the 3D-HST catalogue and derive the spectral slope by fitting the data with a functional form of $F(\lambda) = A\lambda^\beta$. We then use this best-fit function to compute the rest-frame UV flux at 1600\,\AA\, in units of erg s$^{-1}$ cm$^2$ \AA$^{-1}$ and calculate the UV luminosity by using $L_{\rm UV} = 4\pi D_{\rm L}^2F(1600)\lambda_{1600}/(1+z)$, in which $D_{\rm L}$ is the luminosity distance at redshift $z$. The infrared luminosity is adopted from \citet{Swinbank:2014aa}, which uses deblended {\it Herschel} data. The results are plotted in \autoref{IRXb}, in which we also show the composite image of dust, stars, and UV emission. 

Consistent with the trend found by \citet{Casey:2014ab}, with a $L_{\rm IR}$ of 10$^{12.7}$ $L_\odot$ ALESS67.1 lies significantly above the relationships found by other studies for less IR luminous galaxies. As shown in the right panel of \autoref{IRXb} the completely decoupled geometry between rest-frame 280\,$\mu$m and rest-frame 1400\,\AA\, emission confirms the hypothesis that the deviation from the IRX-$\beta$ relationship is caused, at least partially, by the different distributions between dust and UV. For the individual UV-emitting regions, on the other hand, our ALMA data are not sensitive enough to put meaningful upper limits to test whether or not they lie on the relationships (left panel in \autoref{IRXb}). This is the same with the dusty regions where we do not have meaningful constraints on either UV luminosity or slope,

\subsection{The Schmidt-Kennicutt relationship}\label{sec:ks}

\begin{figure*}
	\begin{center}
		\leavevmode
		\includegraphics[scale=1]{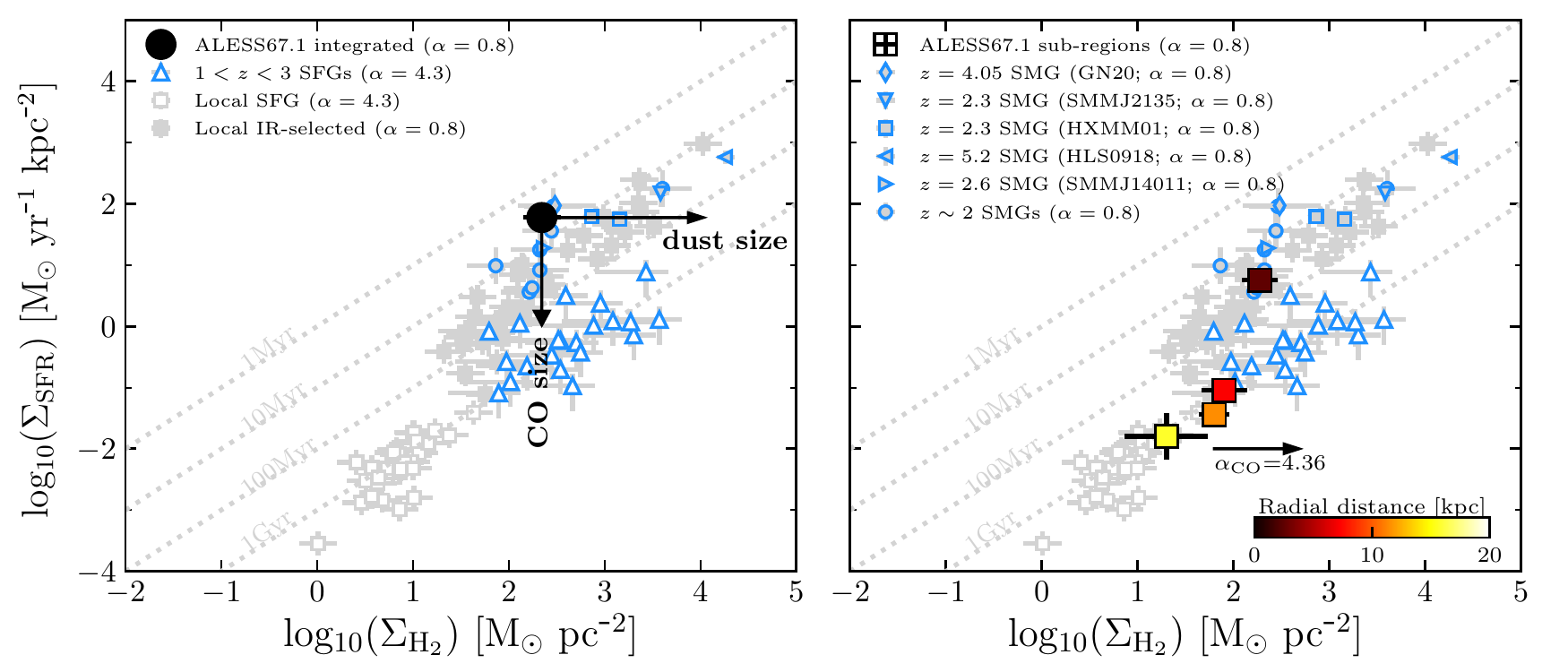}
		\caption{The Schmidt-Kennicut diagram (SFR surface density versus molecular gas surface density), showing ALESS67.1 compared to the literature values on local SFGs, (U)LIRGs \citep{Kennicutt:1998p5718}, $z=1-3$ SFGs \citep{Tacconi:2013aa}, $z\sim2$ SMG \citep{Bothwell:2010p10377,Danielson:2011aa,Fu:2013mm,Sharon:2013aa,Rawle:2014aa}, and a $z\sim4$ SMG (GN20; \citealt{Hodge:2015aa}). The left panel shows ALESS67.1 if we consider the galaxy as a whole, whereas in the right panel the sub-regions (see \autoref{sec:ks} for definition) are shown in squares color coded by the radial distance from the peak of 870\,$\mu$m continuum. In the left panel the downward(rightward) arrows demonstrates where would the data point be if we instead adopt CO(dust) size to measure the SFR(molecular gas) density. In the right panel the rightward arrow indicates the $\sim$0.7 dex shift if we instead adopt galactic $\alpha_{\rm CO}=4.3$. The diagonal dotted lines show the gas consumption time derived by dividing gas density by SFR density. Under the assumption of $\alpha_{\rm CO}=0.8$ the central part of ALESS67.1 ($\lesssim5$\,kpc) agrees with the physical conditions of local (U)LIRGs and other SMGs, whereas the outskirts ($\gtrsim$5\,kpc) of the galaxy follow closely with both local and $z\sim2$ SFGs.
		}
		\label{ks}
	\end{center}
\end{figure*}

The relationship between the surface density of star-formation rate and that of gas has been studied back to the seminal work of \citet{Schmidt:1959aa} and \citet{Kennicutt:1989aa}. Considering galaxies as a whole and assuming a single $\alpha_{\rm CO}$, it was later claimed in observations that this tight correlation, the Schmidt-Kennicutt relationship, with a scatter about a factor of two holds valid in the local star-forming galaxies over a six orders of magnitude in SFR surface density \citep{Kennicutt:1998p5718}. 

However, studies of local interacting mergers have shown a significantly lower $\alpha_{\rm CO}$ ($\alpha_{\rm CO}\sim1$; e.g., \citealt{Scoville:1997aa, Downes:1998aa}), which could also be true at high redshifts based on our analyses of ALESS67.1 in \autoref{sec:alphaCO}. By adopting the lower $\alpha_{\rm CO}$ for (U)LIRGs, together with observations of $z\sim2$ star-forming galaxies and SMGs, it has been claimed that the Schmidt-Kennicutt relationship becomes bimodal, in which IR-luminous galaxies have significantly higher SFR surface density at a fix gas surface density, with a typical gas consumption time ($\Sigma_{\rm gas}/\Sigma_{\rm SFR}$) of $\sim100$\,Myr, in contrast to $\sim$1\,Gyr for normal SFGs (e.g., \citealt{Daddi:2010ab,Genzel:2010aa}). Recent spatially-resolved observations of SMGs on both star-forming regions as traced by dust or radio emissions and CO have found even shorter gas consumption time of $\sim$10\,Myr \citep{Bothwell:2010p10377, Hodge:2015aa}. With measured luminosities and sizes on FIR, CO and H$\alpha$ we are in position to investigate the Schmidt-Kennicutt relationship on a $z\sim2$ merger.

We first compute the global surface density of SFR and H$_2$ for ALESS67.1, by adopting the measured sizes of each tracer and the conversion between $L_{\rm IR}$ and $L_{H\alpha}$ to SFR from \citet{Kennicutt:1998aa}. We assume a Chabrier initial mass function \citep{Chabrier:2003aa} and adopt 0.8 for $\alpha_{\rm CO}$ in order to make comparisons with other studies. The results are plotted in \autoref{ks}, in which we also include literature values from local and high-redshift SFGs and SMGs. We only include measurements obtained from observations that spatially resolve both the star-forming regions and CO, which is the condition of our data.

Our measurements agree with GN20, a SMG at $z\sim4$, as well as some SMGs in the sample of \citet{Bothwell:2010p10377}, suggesting that when looking at the galaxy as a whole some SMGs have a significantly shorter gas consumption time even compared to the local mergers. However, because the dramatic size difference between the dusty star-forming regions, which dominate the SFR surface density, and CO, the comparison of $\Sigma_{\rm H_2}$ and $\Sigma_{\rm SFR}$ on the global scale is non-physical, meaning the two parameters come from regions that are largely unrelated, and therefore the deviation from the typical relationship is expected. To make proper interpretations it is necessary to measure both parameters in resolved regions.

Motivated by the spatial distribution of 870\,$\mu$m continuum, $^{12}$CO($J$=3-2), and H$\alpha$, we construct concentric rings with a width of 0$\farcs$5 ($\sim$4\,kpc) for each ring and centered at the peak of the 870\,$\mu$m continuum. The choice of the width is determined by the spatial resolution of CO. We convolve the 870\,$\mu$m continuum and H$\alpha$ with a CO beam to match the maps in spatial resolution. We then measure $\Sigma_{\rm H_2}$ and $\Sigma_{\rm SFR}$ for each concentric ring and plot the results in the right panel of \autoref{ks}, color coded based on their radial distances. 

Unlike the global values which make ALESS67.1 almost an outlier compared to the local IR-selected galaxies, the measurements in each sub-regions lie in the similar locus occupied by previous measurements. In particular, within one galaxy we obtain a similar bimodal distribution as that seen between normal SFGs and mergers. The transition, which is at $\sim5$\,kpc ($\sim$0.5$''$) in this case, occurs when there is a lack of detection in 870\,$\mu$m, leaving $\Sigma_{\rm SFR}$ derived solely from H$\alpha$. We should stress that, however, these values are derived under the assumption of $\alpha_{\rm CO}=0.8$, and H$\alpha$ is not corrected for the dust extinction, if any. If the sub-regions in the outskirts have a Galactic $\alpha_{\rm CO}$ instead, the derived quantities would shift to higher $\Sigma_{\rm H_2}$ by $\sim$0.7 dex, which would make these regions significantly more gas-rich, or significantly inefficient in star formation. This result is mainly the consequence of the much more extended distribution of $^{12}$CO($J$=3-2) compared to the dust traced by the rest-frame 280\,$\mu$m continuum. 

\section{Summary}\label{sec:summary}
We present detailed spatial and dynamical studies of a SMG, ALESS67.1, using sub-arcsecond resolution ALMA, AO-aided SINFONI, and {\it HST} data to investigate the properties of cold dust (rest-frame 280\,$\mu$m continuum), $^{12}$CO($J$=3-2), optical emission lines (H$\alpha$, [N\,{\sc ii}], [S\,{\sc ii}]), and stellar continuum ({\it HST} imaging). ALESS67.1 has a submillmeter flux ($S_{850}\sim4$\,mJy, SFR$\sim$500\,M$_\odot$ yr$^{-1}$) and redshift ($z=2.12$) which make it typical among the general SMG population that is uncovered in the single-dish submillimeter surveys. Our findings are summarized as:

\begin{enumerate}
	\item By conducting detail dynamical analyses on optical lines and CO, we find that ALESS67.1 is not consistent with an isolated, pure rotating disk. This is supported by the stellar morphology revealed in the {\it HST} imaging, showing tidal features typically seen in major mergers in the local Universe. Considering the compact dusty star formation we conclude that ALESS67.1 is likely a final coalescent-stage merger.
	
	\item We find that the kinematics of H$\alpha$ and $^{12}$CO($J$=3-2) are in broad agreement, although higher SNR CO observations are needed to make a detailed comparison in the central regions.
	
	\item All tracers are resolved at the spatial resolution of our observations, and we have measured half-light radius for each of them, finding 1.2$\pm$0.1\,kpc (circularized) for 870\,$\mu$m (rest-frame 280\,$\mu$m) continuum, 6.5$\pm$0.9\,kpc for $^{12}$CO($J$=3-2), 6.6$\pm$0.9\,kpc for H$\alpha$, 5.1$\pm$1.7\,kpc for [N\,{\sc ii}], 5.1$\pm$2.1\,kpc for [S\,{\sc ii}], and 6.4$\pm$0.5\,kpc for stellar continuum at 1.6\,$\mu$m (rest-frame $\sim$5000\AA). We therefore find that the dust continuum has a factor of 4--6 smaller size than that of strong optical emission lines, NIR continuum, and $^{12}$CO($J$=3-2), and it significantly offsets from the peaks of H$\alpha$ and rest-frame optical stellar continuum. While the $^{12}$CO($J$=3-2) size is consistent with H$\alpha$ and stellar continuum, the peak emissions encloses both obscured (rest-frame FIR continuum) and unobscured (H$\alpha$) star formation, supporting that $^{12}$CO($J$=3-2) traces star-forming gas. 
	
	\item Using the dynamical method we derive the CO-to-H$_2$ conversion factor of $\alpha_{\rm CO} = 1.8\pm1.0$, supported by the estimates based on the gas-to-dust ratio method. Our results are also consistent with the lensed SMGs \citep{Danielson:2011aa, Spilker:2015aa} and the local merger Arp220 \citep{Scoville:1997aa}, and suggest values of $\alpha_{\rm CO}\sim1-2$ are appropriate for high-redshift dusty galaxies.
	
	\item We show that the striking difference in spatial distribution between dust and UV continuum could be part of the reason which drives ALESS67.1, as well as other dusty star-forming galaxies, off of the IRX-$\beta$ relationship found locally, which is widely used to attempt to infer dust-obscured star formation for UV-/optical-selected galaxies at $z\gtrsim4$.
	
	\item We demonstrate that when considering the galaxy as a whole the compact dusty star formation coupled with extended CO could be the cause of unusually high efficiency in star formation found recently in some SMGs (\autoref{ks}). However, when looking at the gas density and SFR density in individual sub-regions (with the assumption of $\alpha_{\rm CO} = 0.8$) we find them consistent with previous studies, in both the core part of the galaxy ($\lesssim5$\,kpc) and the outskirts ($\gtrsim5$\,kpc), although each following different path; The core shares the same locus as mergers whereas the outskirts lie close to the SFGs selected in UV/optical, suggesting different star formation efficiency within one galaxy.

\end{enumerate}

Given the spatial variability of different tracers found within ALESS67.1, our results demonstrate the importance of using high spatial resolution, multi-wavelength data to interpret the properties of SMGs, and more generally also the less IR-luminous $z\gtrsim2$ star-forming galaxies. In particular, there is growing evidence of a geometrical discrepancy between CO and dust among galaxies, regardless of them being SMGs or not, either mismatching in spatial distributions \citep{Riechers:2011aa}, or in sizes \citep{Hodge:2015aa, Spilker:2015aa, Decarli:2016aa}. For less IR-luminous galaxies, cases of mismatch between dust and UV/optical emissions \citep{Koprowski:2016ab}, as well as very extended CO and dust \citep{Ginolfi:2016aa}, have recently been reported. Given now ALMA is reaching its full capability, we expect to see more and more such cases (e.g., \citealt{Tadaki:2017ab}). The physical interpretation of these results, especially the mismatch between CO and dust, needs to be explored further, both theoretically and observationally.  It could be that the rest-frame 280\,$\mu$m continuum still misses a significant amount of cold, optically thin dust, which requires higher surface brightness sensitivity and is better traced by optically-thick CO. Deeper ALMA observations in millimeter wavelengths coupled with hydrodynamical simulations implemented with detailed radiative transfer treatment might shed more light on this issue.

\section{Acknowledgments}
 We acknowledge the referee for a helpful report that has improved the manuscript. We would like to thank Nick Scoville, Desika Narayanan and Gerg\"{o} Popping for discussions. C.-C.C., I.R.S. acknowledge support from the ERC Advanced Investigator programme DUSTYGAL 321334. C.-C.C. also acknowledges the support from the European Southern Observatory through a fellowship program. JAH acknowledges support of the VIDI research programme with project number 639.042.611, which is (partly) financed by the Netherlands Organisation for Scientific Research (NWO). I.R.S. also acknowledges support from a Royal Society/Wolfson Merit Award and STFC through grant number ST/L00075X/1. H.D. acknowledges financial support from the Spanish Ministry of Economy and Competitiveness (MINECO) under the 2014 Ram\'{o}n y Cajal program MINECO RYC-2014-15686. This research made use of Astropy, a community-developed core Python package for Astronomy \citep{Astropy-Collaboration:2013aa}. This research has made use of NASA's Astrophysics Data System.  This paper makes use of the following ALMA data: ADS/JAO.ALMA \# 2012.1.00307.S and 2013.1.00407.S. ALMA is a partnership of ESO (representing its member states), NSF (USA) and NINS (Japan), together with NRC (Canada) and NSC and ASIAA (Taiwan) and KASI (Republic of Korea), in cooperation with the Republic of Chile. The Joint ALMA Observatory is operated by ESO, AUI/NRAO and NAOJ.

\end{CJK}
\bibliography{bib}

\begin{thebibliography}{}
\providecommand\natexlab[1]{#1}
\providecommand\JournalTitle[1]{#1}

\bibitem[{{Alaghband-Zadeh} {et~al.}(2012){Alaghband-Zadeh}, {Chapman},
  {Swinbank}, {Smail}, {Harrison}, {Alexander}, {Casey}, {Dav{\'e}},
  {Narayanan}, {Tamura}, \& {Umehata}}]{Alaghband-Zadeh:2012aa}
{Alaghband-Zadeh}, S., {Chapman}, S.~C., {Swinbank}, A.~M., {et~al.} 2012,
  \href{http://dx.doi.org/10.1111/j.1365-2966.2012.21386.x}{\JournalTitle{\mnras},
  424, 2232}

\bibitem[{{ALMA Partnership} {et~al.}(2015){ALMA Partnership}, {Vlahakis},
  {Hunter}, {Hodge}, {P{\'e}rez}, {Andreani}, {Brogan}, {Cox}, {Martin},
  {Zwaan}, {Matsushita}, {Dent}, {Impellizzeri}, {Fomalont}, {Asaki},
  {Barkats}, {Hills}, {Hirota}, {Kneissl}, {Liuzzo}, {Lucas}, {Marcelino},
  {Nakanishi}, {Phillips}, {Richards}, {Toledo}, {Aladro}, {Broguiere},
  {Cortes}, {Cortes}, {Espada}, {Galarza}, {Garcia-Appadoo}, {Guzman-Ramirez},
  {Hales}, {Humphreys}, {Jung}, {Kameno}, {Laing}, {Leon}, {Marconi},
  {Mignano}, {Nikolic}, {Nyman}, {Radiszcz}, {Remijan}, {Rod{\'o}n}, {Sawada},
  {Takahashi}, {Tilanus}, {Vila Vilaro}, {Watson}, {Wiklind}, {Ao}, {Di
  Francesco}, {Hatsukade}, {Hatziminaoglou}, {Mangum}, {Matsuda}, {van Kampen},
  {Wootten}, {de Gregorio-Monsalvo}, {Dumas}, {Francke}, {Gallardo}, {Garcia},
  {Gonzalez}, {Hill}, {Iono}, {Kaminski}, {Karim}, {Krips}, {Kurono},
  {Lonsdale}, {Lopez}, {Morales}, {Plarre}, {Videla}, {Villard}, {Hibbard}, \&
  {Tatematsu}}]{ALMA-Partnership:2015aa}
{ALMA Partnership}, {Vlahakis}, C., {Hunter}, T.~R., {et~al.} 2015,
  \href{http://dx.doi.org/10.1088/2041-8205/808/1/L4}{\JournalTitle{\apjl},
  808, L4}

\bibitem[{{Astropy Collaboration} {et~al.}(2013){Astropy Collaboration},
  {Robitaille}, {Tollerud}, {Greenfield}, {Droettboom}, {Bray}, {Aldcroft},
  {Davis}, {Ginsburg}, {Price-Whelan}, {Kerzendorf}, {Conley}, {Crighton},
  {Barbary}, {Muna}, {Ferguson}, {Grollier}, {Parikh}, {Nair}, {Unther},
  {Deil}, {Woillez}, {Conseil}, {Kramer}, {Turner}, {Singer}, {Fox}, {Weaver},
  {Zabalza}, {Edwards}, {Azalee Bostroem}, {Burke}, {Casey}, {Crawford},
  {Dencheva}, {Ely}, {Jenness}, {Labrie}, {Lim}, {Pierfederici}, {Pontzen},
  {Ptak}, {Refsdal}, {Servillat}, \&
  {Streicher}}]{Astropy-Collaboration:2013aa}
{Astropy Collaboration}, {Robitaille}, T.~P., {Tollerud}, E.~J., {et~al.} 2013,
  \href{http://dx.doi.org/10.1051/0004-6361/201322068}{\JournalTitle{\aap},
  558, A33}

\bibitem[{{Bacon} {et~al.}(2001){Bacon}, {Copin}, {Monnet}, {Miller},
  {Allington-Smith}, {Bureau}, {Carollo}, {Davies}, {Emsellem}, {Kuntschner},
  {Peletier}, {Verolme}, \& {de Zeeuw}}]{Bacon:2001aa}
{Bacon}, R., {Copin}, Y., {Monnet}, G., {et~al.} 2001,
  \href{http://dx.doi.org/10.1046/j.1365-8711.2001.04612.x}{\JournalTitle{\mnras},
  326, 23}

\bibitem[{{Barger} {et~al.}(1998){Barger}, {Cowie}, {Sanders}, {Fulton},
  {Taniguchi}, {Sato}, {Kawara}, \& {Okuda}}]{Barger:1998p13566}
{Barger}, A.~J., {Cowie}, L.~L., {Sanders}, D.~B., {et~al.} 1998,
  \href{http://dx.doi.org/10.1038/28338}{\JournalTitle{\nat}, 394, 248}

\bibitem[{{Barger} {et~al.}(2012){Barger}, {Wang}, {Cowie}, {Owen}, {Chen}, \&
  {Williams}}]{Barger:2012lr}
{Barger}, A.~J., {Wang}, W.-H., {Cowie}, L.~L., {et~al.} 2012,
  \href{http://dx.doi.org/10.1088/0004-637X/761/2/89}{\JournalTitle{\apj}, 761,
  89}

\bibitem[{{Barger} {et~al.}(2014){Barger}, {Cowie}, {Chen}, {Owen}, {Wang},
  {Casey}, {Lee}, {Sanders}, \& {Williams}}]{Barger:2014aa}
{Barger}, A.~J., {Cowie}, L.~L., {Chen}, C.-C., {et~al.} 2014,
  \href{http://dx.doi.org/10.1088/0004-637X/784/1/9}{\JournalTitle{\apj}, 784,
  9}

\bibitem[{{Baugh} {et~al.}(2005){Baugh}, {Lacey}, {Frenk}, {Granato}, {Silva},
  {Bressan}, {Benson}, \& {Cole}}]{Baugh:2005p14519}
{Baugh}, C.~M., {Lacey}, C.~G., {Frenk}, C.~S., {et~al.} 2005,
  \href{http://dx.doi.org/10.1111/j.1365-2966.2004.08553.x}{\JournalTitle{\mnras},
  356, 1191}

\bibitem[{{Begeman}(1989)}]{Begeman:1989aa}
{Begeman}, K.~G. 1989, \JournalTitle{\aap}, 223, 47

\bibitem[{{Bolatto} {et~al.}(2013){Bolatto}, {Wolfire}, \&
  {Leroy}}]{Bolatto:2013aa}
{Bolatto}, A.~D., {Wolfire}, M., \& {Leroy}, A.~K. 2013,
  \href{http://dx.doi.org/10.1146/annurev-astro-082812-140944}{\JournalTitle{\araa},
  51, 207}

\bibitem[{{Bothwell} {et~al.}(2010){Bothwell}, {Chapman}, {Tacconi}, {Smail},
  {Ivison}, {Casey}, {Bertoldi}, {Beswick}, {Biggs}, {Blain}, {Cox}, {Genzel},
  {Greve}, {Kennicutt}, {Muxlow}, {Neri}, \& {Omont}}]{Bothwell:2010p10377}
{Bothwell}, M.~S., {Chapman}, S.~C., {Tacconi}, L., {et~al.} 2010,
  \href{http://dx.doi.org/10.1111/j.1365-2966.2010.16480.x}{\JournalTitle{\mnras},
  405, 219}

\bibitem[{{Bothwell} {et~al.}(2013){Bothwell}, {Smail}, {Chapman}, {Genzel},
  {Ivison}, {Tacconi}, {Alaghband-Zadeh}, {Bertoldi}, {Blain}, {Casey}, {Cox},
  {Greve}, {Lutz}, {Neri}, {Omont}, \& {Swinbank}}]{Bothwell:2013lp}
{Bothwell}, M.~S., {Smail}, I., {Chapman}, S.~C., {et~al.} 2013,
  \JournalTitle{\mnras}, 429, 3047

\bibitem[{{Bouch{\'e}} {et~al.}(2007){Bouch{\'e}}, {Cresci}, {Davies},
  {Eisenhauer}, {F{\"o}rster Schreiber}, {Genzel}, {Gillessen}, {Lehnert},
  {Lutz}, {Nesvadba}, {Shapiro}, {Sternberg}, {Tacconi}, {Verma}, {Cimatti},
  {Daddi}, {Renzini}, {Erb}, {Shapley}, \& {Steidel}}]{Bouche:2007aa}
{Bouch{\'e}}, N., {Cresci}, G., {Davies}, R., {et~al.} 2007,
  \href{http://dx.doi.org/10.1086/522221}{\JournalTitle{\apj}, 671, 303}

\bibitem[{{Bournaud} {et~al.}(2014){Bournaud}, {Perret}, {Renaud}, {Dekel},
  {Elmegreen}, {Elmegreen}, {Teyssier}, {Amram}, {Daddi}, {Duc}, {Elbaz},
  {Epinat}, {Gabor}, {Juneau}, {Kraljic}, \& {Le Floch'}}]{Bournaud:2014aa}
{Bournaud}, F., {Perret}, V., {Renaud}, F., {et~al.} 2014,
  \href{http://dx.doi.org/10.1088/0004-637X/780/1/57}{\JournalTitle{\apj}, 780,
  57}

\bibitem[{{Bouwens} {et~al.}(2015){Bouwens}, {Illingworth}, {Oesch}, {Trenti},
  {Labb{\'e}}, {Bradley}, {Carollo}, {van Dokkum}, {Gonzalez}, {Holwerda},
  {Franx}, {Spitler}, {Smit}, \& {Magee}}]{Bouwens:2015aa}
{Bouwens}, R.~J., {Illingworth}, G.~D., {Oesch}, P.~A., {et~al.} 2015,
  \href{http://dx.doi.org/10.1088/0004-637X/803/1/34}{\JournalTitle{\apj}, 803,
  34}

\bibitem[{{Brammer} {et~al.}(2012){Brammer}, {van Dokkum}, {Franx},
  {Fumagalli}, {Patel}, {Rix}, {Skelton}, {Kriek}, {Nelson}, {Schmidt},
  {Bezanson}, {da Cunha}, {Erb}, {Fan}, {F{\"o}rster Schreiber}, {Illingworth},
  {Labb{\'e}}, {Leja}, {Lundgren}, {Magee}, {Marchesini}, {McCarthy},
  {Momcheva}, {Muzzin}, {Quadri}, {Steidel}, {Tal}, {Wake}, {Whitaker}, \&
  {Williams}}]{Brammer:2012aa}
{Brammer}, G.~B., {van Dokkum}, P.~G., {Franx}, M., {et~al.} 2012,
  \href{http://dx.doi.org/10.1088/0067-0049/200/2/13}{\JournalTitle{\apjs},
  200, 13}

\bibitem[{{Buat} {et~al.}(2005){Buat}, {Iglesias-P{\'a}ramo}, {Seibert},
  {Burgarella}, {Charlot}, {Martin}, {Xu}, {Heckman}, {Boissier}, {Boselli},
  {Barlow}, {Bianchi}, {Byun}, {Donas}, {Forster}, {Friedman}, {Jelinski},
  {Lee}, {Madore}, {Malina}, {Milliard}, {Morissey}, {Neff}, {Rich},
  {Schiminovitch}, {Siegmund}, {Small}, {Szalay}, {Welsh}, \&
  {Wyder}}]{Buat:2005aa}
{Buat}, V., {Iglesias-P{\'a}ramo}, J., {Seibert}, M., {et~al.} 2005,
  \href{http://dx.doi.org/10.1086/423241}{\JournalTitle{\apjl}, 619, L51}

\bibitem[{{Burkert} {et~al.}(2010){Burkert}, {Genzel}, {Bouch{\'e}}, {Cresci},
  {Khochfar}, {Sommer-Larsen}, {Sternberg}, {Naab}, {F{\"o}rster Schreiber},
  {Tacconi}, {Shapiro}, {Hicks}, {Lutz}, {Davies}, {Buschkamp}, \&
  {Genel}}]{Burkert:2010aa}
{Burkert}, A., {Genzel}, R., {Bouch{\'e}}, N., {et~al.} 2010,
  \href{http://dx.doi.org/10.1088/0004-637X/725/2/2324}{\JournalTitle{\apj},
  725, 2324}

\bibitem[{{Calanog} {et~al.}(2014){Calanog}, {Fu}, {Cooray}, {Wardlow}, {Ma},
  {Amber}, {Baker}, {Baes}, {Bock}, {Bourne}, {Bussmann}, {Casey}, {Chapman},
  {Clements}, {Conley}, {Dannerbauer}, {De Zotti}, {Dunne}, {Dye}, {Eales},
  {Farrah}, {Furlanetto}, {Harris}, {Ivison}, {Kim}, {Maddox}, {Magdis},
  {Messias}, {Micha{\l}owski}, {Negrello}, {Nightingale}, {O'Bryan}, {Oliver},
  {Riechers}, {Scott}, {Serjeant}, {Simpson}, {Smith}, {Timmons}, {Thacker},
  {Valiante}, \& {Vieira}}]{Calanog:2014aa}
{Calanog}, J.~A., {Fu}, H., {Cooray}, A., {et~al.} 2014,
  \href{http://dx.doi.org/10.1088/0004-637X/797/2/138}{\JournalTitle{\apj},
  797, 138}

\bibitem[{{Casey} {et~al.}(2014{\natexlab{a}}){Casey}, {Narayanan}, \&
  {Cooray}}]{Casey:2014aa}
{Casey}, C.~M., {Narayanan}, D., \& {Cooray}, A. 2014{\natexlab{a}},
  \href{http://dx.doi.org/10.1016/j.physrep.2014.02.009}{\JournalTitle{\physrep},
  541, 45}

\bibitem[{{Casey} {et~al.}(2014{\natexlab{b}}){Casey}, {Scoville}, {Sanders},
  {Lee}, {Cooray}, {Finkelstein}, {Capak}, {Conley}, {De Zotti}, {Farrah},
  {Fu}, {Le Floc'h}, {Ilbert}, {Ivison}, \& {Takeuchi}}]{Casey:2014ab}
{Casey}, C.~M., {Scoville}, N.~Z., {Sanders}, D.~B., {et~al.}
  2014{\natexlab{b}},
  \href{http://dx.doi.org/10.1088/0004-637X/796/2/95}{\JournalTitle{\apj}, 796,
  95}

\bibitem[{{Chabrier}(2003)}]{Chabrier:2003aa}
{Chabrier}, G. 2003,
  \href{http://dx.doi.org/10.1086/376392}{\JournalTitle{\pasp}, 115, 763}

\bibitem[{{Chapman} {et~al.}(2005){Chapman}, {Blain}, {Smail}, \&
  {Ivison}}]{Chapman:2005p5778}
{Chapman}, S.~C., {Blain}, A.~W., {Smail}, I., \& {Ivison}, R.~J. 2005,
  \href{http://dx.doi.org/10.1086/428082}{\JournalTitle{\apj}, 622, 772}

\bibitem[{{Chen} {et~al.}(2015){Chen}, {Smail}, {Swinbank}, {Simpson}, {Ma},
  {Alexander}, {Biggs}, {Brandt}, {Chapman}, {Coppin}, {Danielson},
  {Dannerbauer}, {Edge}, {Greve}, {Ivison}, {Karim}, {Menten}, {Schinnerer},
  {Walter}, {Wardlow}, {Wei{\ss}}, \& {van der Werf}}]{Chen:2015aa}
{Chen}, C.-C., {Smail}, I., {Swinbank}, A.~M., {et~al.} 2015,
  \href{http://dx.doi.org/10.1088/0004-637X/799/2/194}{\JournalTitle{\apj},
  799, 194}

\bibitem[{{Chen} {et~al.}(2016){Chen}, {Smail}, {Ivison}, {Arumugam},
  {Almaini}, {Conselice}, {Geach}, {Hartley}, {Ma}, {Mortlock}, {Simpson},
  {Simpson}, {Swinbank}, {Aretxaga}, {Blain}, {Chapman}, {Dunlop}, {Farrah},
  {Halpern}, {Micha{\l}owski}, {van der Werf}, {Wilkinson}, \&
  {Zavala}}]{Chen:2016aa}
{Chen}, C.-C., {Smail}, I., {Ivison}, R.~J., {et~al.} 2016,
  \href{http://dx.doi.org/10.3847/0004-637X/820/2/82}{\JournalTitle{\apj}, 820,
  82}

\bibitem[{{Courteau}(1997)}]{Courteau:1997aa}
{Courteau}, S. 1997,
  \href{http://dx.doi.org/10.1086/118656}{\JournalTitle{\aj}, 114, 2402}

\bibitem[{{Cowley} {et~al.}(2015){Cowley}, {Lacey}, {Baugh}, \&
  {Cole}}]{Cowley:2015aa}
{Cowley}, W.~I., {Lacey}, C.~G., {Baugh}, C.~M., \& {Cole}, S. 2015,
  \href{http://dx.doi.org/10.1093/mnras/stu2179}{\JournalTitle{\mnras}, 446,
  1784}

\bibitem[{{da Cunha} {et~al.}(2015){da Cunha}, {Walter}, {Smail}, {Swinbank},
  {Simpson}, {Decarli}, {Hodge}, {Weiss}, {van der Werf}, {Bertoldi},
  {Chapman}, {Cox}, {Danielson}, {Dannerbauer}, {Greve}, {Ivison}, {Karim}, \&
  {Thomson}}]{da-Cunha:2015aa}
{da Cunha}, E., {Walter}, F., {Smail}, I.~R., {et~al.} 2015,
  \href{http://dx.doi.org/10.1088/0004-637X/806/1/110}{\JournalTitle{\apj},
  806, 110}

\bibitem[{{Daddi} {et~al.}(2010){Daddi}, {Elbaz}, {Walter}, {Bournaud},
  {Salmi}, {Carilli}, {Dannerbauer}, {Dickinson}, {Monaco}, \&
  {Riechers}}]{Daddi:2010ab}
{Daddi}, E., {Elbaz}, D., {Walter}, F., {et~al.} 2010,
  \href{http://dx.doi.org/10.1088/2041-8205/714/1/L118}{\JournalTitle{\apjl},
  714, L118}

\bibitem[{{Danielson} {et~al.}(2011){Danielson}, {Swinbank}, {Smail}, {Cox},
  {Edge}, {Weiss}, {Harris}, {Baker}, {De Breuck}, {Geach}, {Ivison}, {Krips},
  {Lundgren}, {Longmore}, {Neri}, \& {Flaquer}}]{Danielson:2011aa}
{Danielson}, A.~L.~R., {Swinbank}, A.~M., {Smail}, I., {et~al.} 2011,
  \href{http://dx.doi.org/10.1111/j.1365-2966.2010.17549.x}{\JournalTitle{\mnras},
  410, 1687}

\bibitem[{{Danielson} {et~al.}(2013){Danielson}, {Swinbank}, {Smail}, {Bayet},
  {van der Werf}, {Cox}, {Edge}, {Henkel}, \& {Ivison}}]{Danielson:2013aa}
---. 2013,
  \href{http://dx.doi.org/10.1093/mnras/stt1775}{\JournalTitle{\mnras}, 436,
  2793}

\bibitem[{{Danielson} {et~al.}(2017){Danielson}, {Swinbank}, {Smail},
  {Simpson}, {Casey}, {Chapman}, {da Cunha}, {Hodge}, {Walter}, {Wardlow},
  {Alexander}, {Brandt}, {de Breuck}, {Coppin}, {Dannerbauer}, {Dickinson},
  {Edge}, {Gawiser}, {Ivison}, {Karim}, {Kovacs}, {Lutz}, {Menten},
  {Schinnerer}, {Wei{\ss}}, \& {van der Werf}}]{Danielson:2017aa}
---. 2017,
  \href{http://dx.doi.org/10.3847/1538-4357/aa6caf}{\JournalTitle{\apj}, 840,
  78}

\bibitem[{{Dav{\'e}} {et~al.}(2010){Dav{\'e}}, {Finlator}, {Oppenheimer},
  {Fardal}, {Katz}, {Kere{\v s}}, \& {Weinberg}}]{Dave:2010kx}
{Dav{\'e}}, R., {Finlator}, K., {Oppenheimer}, B.~D., {et~al.} 2010,
  \href{http://dx.doi.org/10.1111/j.1365-2966.2010.16395.x}{\JournalTitle{\mnras},
  404, 1355}

\bibitem[{{De Breuck} {et~al.}(2014){De Breuck}, {Williams}, {Swinbank},
  {Caselli}, {Coppin}, {Davis}, {Maiolino}, {Nagao}, {Smail}, {Walter},
  {Wei{\ss}}, \& {Zwaan}}]{De-Breuck:2014aa}
{De Breuck}, C., {Williams}, R.~J., {Swinbank}, M., {et~al.} 2014,
  \href{http://dx.doi.org/10.1051/0004-6361/201323331}{\JournalTitle{\aap},
  565, A59}

\bibitem[{{Decarli} {et~al.}(2016){Decarli}, {Walter}, {Aravena}, {Carilli},
  {Bouwens}, {da Cunha}, {Daddi}, {Elbaz}, {Riechers}, {Smail}, {Swinbank},
  {Weiss}, {Bacon}, {Bauer}, {Bell}, {Bertoldi}, {Chapman}, {Colina}, {Cortes},
  {Cox}, {G{\'o}nzalez-L{\'o}pez}, {Inami}, {Ivison}, {Hodge}, {Karim},
  {Magnelli}, {Ota}, {Popping}, {Rix}, {Sargent}, {van der Wel}, \& {van der
  Werf}}]{Decarli:2016aa}
{Decarli}, R., {Walter}, F., {Aravena}, M., {et~al.} 2016,
  \href{http://dx.doi.org/10.3847/1538-4357/833/1/70}{\JournalTitle{\apj}, 833,
  70}

\bibitem[{{Downes} \& {Solomon}(1998)}]{Downes:1998aa}
{Downes}, D., \& {Solomon}, P.~M. 1998,
  \href{http://dx.doi.org/10.1086/306339}{\JournalTitle{\apj}, 507, 615}

\bibitem[{{Engel} {et~al.}(2010){Engel}, {Tacconi}, {Davies}, {Neri}, {Smail},
  {Chapman}, {Genzel}, {Cox}, {Greve}, {Ivison}, {Blain}, {Bertoldi}, \&
  {Omont}}]{Engel:2010p9470}
{Engel}, H., {Tacconi}, L.~J., {Davies}, R.~I., {et~al.} 2010,
  \href{http://dx.doi.org/10.1088/0004-637X/724/1/233}{\JournalTitle{\apj},
  724, 233}

\bibitem[{{Foreman-Mackey} {et~al.}(2013){Foreman-Mackey}, {Hogg}, {Lang}, \&
  {Goodman}}]{Foreman-Mackey:2013aa}
{Foreman-Mackey}, D., {Hogg}, D.~W., {Lang}, D., \& {Goodman}, J. 2013,
  \href{http://dx.doi.org/10.1086/670067}{\JournalTitle{\pasp}, 125, 306}

\bibitem[{{F{\"o}rster Schreiber} {et~al.}(2009){F{\"o}rster Schreiber},
  {Genzel}, {Bouch{\'e}}, {Cresci}, {Davies}, {Buschkamp}, {Shapiro},
  {Tacconi}, {Hicks}, {Genel}, {Shapley}, {Erb}, {Steidel}, {Lutz},
  {Eisenhauer}, {Gillessen}, {Sternberg}, {Renzini}, {Cimatti}, {Daddi},
  {Kurk}, {Lilly}, {Kong}, {Lehnert}, {Nesvadba}, {Verma}, {McCracken},
  {Arimoto}, {Mignoli}, \& {Onodera}}]{Forster-Schreiber:2009aa}
{F{\"o}rster Schreiber}, N.~M., {Genzel}, R., {Bouch{\'e}}, N., {et~al.} 2009,
  \href{http://dx.doi.org/10.1088/0004-637X/706/2/1364}{\JournalTitle{\apj},
  706, 1364}

\bibitem[{{Fu} {et~al.}(2013){Fu}, {Cooray}, {Feruglio}, {Ivison}, {Riechers},
  {Gurwell}, {Bussmann}, {Harris}, {Altieri}, {Aussel}, {Baker}, {Bock},
  {Boylan-Kolchin}, {Bridge}, {Calanog}, {Casey}, {Cava}, {Chapman},
  {Clements}, {Conley}, {Cox}, {Farrah}, {Frayer}, {Hopwood}, {Jia}, {Magdis},
  {Marsden}, {Mart{\'{\i}}nez-Navajas}, {Negrello}, {Neri}, {Oliver}, {Omont},
  {Page}, {P{\'e}rez-Fournon}, {Schulz}, {Scott}, {Smith}, {Vaccari},
  {Valtchanov}, {Vieira}, {Viero}, {Wang}, {Wardlow}, \& {Zemcov}}]{Fu:2013mm}
{Fu}, H., {Cooray}, A., {Feruglio}, C., {et~al.} 2013,
  \href{http://dx.doi.org/10.1038/nature12184}{\JournalTitle{\nat}, 498, 338}

\bibitem[{{Genzel} {et~al.}(2010){Genzel}, {Tacconi}, {Gracia-Carpio},
  {Sternberg}, {Cooper}, {Shapiro}, {Bolatto}, {Bouch{\'e}}, {Bournaud},
  {Burkert}, {Combes}, {Comerford}, {Cox}, {Davis}, {Schreiber},
  {Garcia-Burillo}, {Lutz}, {Naab}, {Neri}, {Omont}, {Shapley}, \&
  {Weiner}}]{Genzel:2010aa}
{Genzel}, R., {Tacconi}, L.~J., {Gracia-Carpio}, J., {et~al.} 2010,
  \href{http://dx.doi.org/10.1111/j.1365-2966.2010.16969.x}{\JournalTitle{\mnras},
  407, 2091}

\bibitem[{{Genzel} {et~al.}(2017){Genzel}, {Schreiber}, {{\"U}bler}, {Lang},
  {Naab}, {Bender}, {Tacconi}, {Wisnioski}, {Wuyts}, {Alexander}, {Beifiori},
  {Belli}, {Brammer}, {Burkert}, {Carollo}, {Chan}, {Davies}, {Fossati},
  {Galametz}, {Genel}, {Gerhard}, {Lutz}, {Mendel}, {Momcheva}, {Nelson},
  {Renzini}, {Saglia}, {Sternberg}, {Tacchella}, {Tadaki}, \&
  {Wilman}}]{Genzel:2017aa}
{Genzel}, R., {Schreiber}, N.~M.~F., {{\"U}bler}, H., {et~al.} 2017,
  \href{http://dx.doi.org/10.1038/nature21685}{\JournalTitle{\nat}, 543, 397}

\bibitem[{{Gil de Paz} {et~al.}(2007){Gil de Paz}, {Boissier}, {Madore},
  {Seibert}, {Joe}, {Boselli}, {Wyder}, {Thilker}, {Bianchi}, {Rey}, {Rich},
  {Barlow}, {Conrow}, {Forster}, {Friedman}, {Martin}, {Morrissey}, {Neff},
  {Schiminovich}, {Small}, {Donas}, {Heckman}, {Lee}, {Milliard}, {Szalay}, \&
  {Yi}}]{Gil-de-Paz:2007aa}
{Gil de Paz}, A., {Boissier}, S., {Madore}, B.~F., {et~al.} 2007,
  \href{http://dx.doi.org/10.1086/516636}{\JournalTitle{\apjs}, 173, 185}

\bibitem[{{Ginolfi} {et~al.}(2016){Ginolfi}, {Maiolino}, {Nagao}, {Carniani},
  {Belfiore}, {Cresci}, {Hatsukade}, {Mannucci}, {Marconi}, {Pallottini},
  {Schneider}, \& {Santini}}]{Ginolfi:2016aa}
{Ginolfi}, M., {Maiolino}, R., {Nagao}, T., {et~al.} 2016, \JournalTitle{MNRAS
  in press}, \href{http://arxiv.org/abs/1611.07026}{{\sffamily
  arXiv:1611.07026}}

\bibitem[{{Greisen} {et~al.}(2006){Greisen}, {Calabretta}, {Valdes}, \&
  {Allen}}]{Greisen:2006aa}
{Greisen}, E.~W., {Calabretta}, M.~R., {Valdes}, F.~G., \& {Allen}, S.~L. 2006,
  \href{http://dx.doi.org/10.1051/0004-6361:20053818}{\JournalTitle{\aap}, 446,
  747}

\bibitem[{{Grogin} {et~al.}(2011){Grogin}, {Kocevski}, {Faber}, {Ferguson},
  {Koekemoer}, {Riess}, {Acquaviva}, {Alexander}, {Almaini}, {Ashby}, {Barden},
  {Bell}, {Bournaud}, {Brown}, {Caputi}, {Casertano}, {Cassata}, {Castellano},
  {Challis}, {Chary}, {Cheung}, {Cirasuolo}, {Conselice}, {Roshan Cooray},
  {Croton}, {Daddi}, {Dahlen}, {Dav{\'e}}, {de Mello}, {Dekel}, {Dickinson},
  {Dolch}, {Donley}, {Dunlop}, {Dutton}, {Elbaz}, {Fazio}, {Filippenko},
  {Finkelstein}, {Fontana}, {Gardner}, {Garnavich}, {Gawiser}, {Giavalisco},
  {Grazian}, {Guo}, {Hathi}, {H{\"a}ussler}, {Hopkins}, {Huang}, {Huang},
  {Jha}, {Kartaltepe}, {Kirshner}, {Koo}, {Lai}, {Lee}, {Li}, {Lotz}, {Lucas},
  {Madau}, {McCarthy}, {McGrath}, {McIntosh}, {McLure}, {Mobasher},
  {Moustakas}, {Mozena}, {Nandra}, {Newman}, {Niemi}, {Noeske}, {Papovich},
  {Pentericci}, {Pope}, {Primack}, {Rajan}, {Ravindranath}, {Reddy}, {Renzini},
  {Rix}, {Robaina}, {Rodney}, {Rosario}, {Rosati}, {Salimbeni}, {Scarlata},
  {Siana}, {Simard}, {Smidt}, {Somerville}, {Spinrad}, {Straughn}, {Strolger},
  {Telford}, {Teplitz}, {Trump}, {van der Wel}, {Villforth}, {Wechsler},
  {Weiner}, {Wiklind}, {Wild}, {Wilson}, {Wuyts}, {Yan}, \&
  {Yun}}]{Grogin:2011fj}
{Grogin}, N.~A., {Kocevski}, D.~D., {Faber}, S.~M., {et~al.} 2011,
  \href{http://dx.doi.org/10.1088/0067-0049/197/2/35}{\JournalTitle{\apjs},
  197, 35}

\bibitem[{{Gunn} {et~al.}(2006){Gunn}, {Siegmund}, {Mannery}, {Owen}, {Hull},
  {Leger}, {Carey}, {Knapp}, {York}, {Boroski}, {Kent}, {Lupton}, {Rockosi},
  {Evans}, {Waddell}, {Anderson}, {Annis}, {Barentine}, {Bartoszek}, {Bastian},
  {Bracker}, {Brewington}, {Briegel}, {Brinkmann}, {Brown}, {Carr},
  {Czarapata}, {Drennan}, {Dombeck}, {Federwitz}, {Gillespie}, {Gonzales},
  {Hansen}, {Harvanek}, {Hayes}, {Jordan}, {Kinney}, {Klaene}, {Kleinman},
  {Kron}, {Kresinski}, {Lee}, {Limmongkol}, {Lindenmeyer}, {Long}, {Loomis},
  {McGehee}, {Mantsch}, {Neilsen}, {Neswold}, {Newman}, {Nitta}, {Peoples},
  {Pier}, {Prieto}, {Prosapio}, {Rivetta}, {Schneider}, {Snedden}, \&
  {Wang}}]{Gunn:2006aa}
{Gunn}, J.~E., {Siegmund}, W.~A., {Mannery}, E.~J., {et~al.} 2006,
  \href{http://dx.doi.org/10.1086/500975}{\JournalTitle{\aj}, 131, 2332}

\bibitem[{{Harris} {et~al.}(2010){Harris}, {Baker}, {Zonak}, {Sharon},
  {Genzel}, {Rauch}, {Watts}, \& {Creager}}]{Harris:2010p11118}
{Harris}, A.~I., {Baker}, A.~J., {Zonak}, S.~G., {et~al.} 2010,
  \href{http://dx.doi.org/10.1088/0004-637X/723/2/1139}{\JournalTitle{\apj},
  723, 1139}

\bibitem[{{Harrison} {et~al.}(2012){Harrison}, {Alexander}, {Swinbank},
  {Smail}, {Alaghband-Zadeh}, {Bauer}, {Chapman}, {Del Moro}, {Hickox},
  {Ivison}, {Men{\'e}ndez-Delmestre}, {Mullaney}, \&
  {Nesvadba}}]{Harrison:2012aa}
{Harrison}, C.~M., {Alexander}, D.~M., {Swinbank}, A.~M., {et~al.} 2012,
  \href{http://dx.doi.org/10.1111/j.1365-2966.2012.21723.x}{\JournalTitle{\mnras},
  426, 1073}

\bibitem[{{Hayward} {et~al.}(2013){Hayward}, {Narayanan}, {Kere{\v s}},
  {Hayward}, {Narayanan}, {Kere{\v s}}, {Jonsson}, {Hopkins}, {Cox}, \&
  {Hernquist}}]{Hayward:2013lr}
{Hayward}, C.~C., {Narayanan}, D., {Kere{\v s}}, D., {et~al.} 2013,
  \JournalTitle{\mnras}, 428, 2529

\bibitem[{{Hodge} {et~al.}(2012){Hodge}, {Carilli}, {Walter}, {de Blok},
  {Riechers}, {Daddi}, \& {Lentati}}]{Hodge:2012fk}
{Hodge}, J.~A., {Carilli}, C.~L., {Walter}, F., {et~al.} 2012,
  \href{http://dx.doi.org/10.1088/0004-637X/760/1/11}{\JournalTitle{\apj}, 760,
  11}

\bibitem[{{Hodge} {et~al.}(2015){Hodge}, {Riechers}, {Decarli}, {Walter},
  {Carilli}, {Daddi}, \& {Dannerbauer}}]{Hodge:2015aa}
{Hodge}, J.~A., {Riechers}, D., {Decarli}, R., {et~al.} 2015,
  \href{http://dx.doi.org/10.1088/2041-8205/798/1/L18}{\JournalTitle{\apjl},
  798, L18}

\bibitem[{{Hodge} {et~al.}(2013){Hodge}, {Karim}, {Smail}, {Swinbank},
  {Walter}, {Biggs}, {Ivison}, {Weiss}, {Alexander}, {Bertoldi}, {Brandt},
  {Chapman}, {Coppin}, {Cox}, {Danielson}, {Dannerbauer}, {De Breuck},
  {Decarli}, {Edge}, {Greve}, {Knudsen}, {Menten}, {Rix}, {Schinnerer},
  {Simpson}, {Wardlow}, \& {van der Werf}}]{Hodge:2013lr}
{Hodge}, J.~A., {Karim}, A., {Smail}, I., {et~al.} 2013,
  \href{http://dx.doi.org/10.1088/0004-637X/768/1/91}{\JournalTitle{\apj}, 768,
  91}

\bibitem[{{Hodge} {et~al.}(2016){Hodge}, {Swinbank}, {Simpson}, {Smail},
  {Walter}, {Alexander}, {Bertoldi}, {Biggs}, {Brandt}, {Chapman}, {Chen},
  {Coppin}, {Cox}, {Dannerbauer}, {Edge}, {Greve}, {Ivison}, {Karim},
  {Knudsen}, {Menten}, {Rix}, {Schinnerer}, {Wardlow}, {Weiss}, \& {van der
  Werf}}]{Hodge:2016aa}
{Hodge}, J.~A., {Swinbank}, A.~M., {Simpson}, J.~M., {et~al.} 2016,
  \href{http://dx.doi.org/10.3847/1538-4357/833/1/103}{\JournalTitle{\apj},
  833, 103}

\bibitem[{{Hopkins} {et~al.}(2013){Hopkins}, {Cox}, {Hernquist}, {Narayanan},
  {Hayward}, \& {Murray}}]{Hopkins:2013aa}
{Hopkins}, P.~F., {Cox}, T.~J., {Hernquist}, L., {et~al.} 2013,
  \href{http://dx.doi.org/10.1093/mnras/stt017}{\JournalTitle{\mnras}, 430,
  1901}

\bibitem[{{Howell} {et~al.}(2010){Howell}, {Armus}, {Mazzarella}, {Evans},
  {Surace}, {Sanders}, {Petric}, {Appleton}, {Bothun}, {Bridge}, {Chan},
  {Charmandaris}, {Frayer}, {Haan}, {Inami}, {Kim}, {Lord}, {Madore},
  {Melbourne}, {Schulz}, {U}, {Vavilkin}, {Veilleux}, \& {Xu}}]{Howell:2010aa}
{Howell}, J.~H., {Armus}, L., {Mazzarella}, J.~M., {et~al.} 2010,
  \href{http://dx.doi.org/10.1088/0004-637X/715/1/572}{\JournalTitle{\apj},
  715, 572}

\bibitem[{{Hughes} {et~al.}(1998){Hughes}, {Serjeant}, {Dunlop},
  {Rowan-Robinson}, {Blain}, {Mann}, {Ivison}, {Peacock}, {Efstathiou}, {Gear},
  {Oliver}, {Lawrence}, {Longair}, {Goldschmidt}, \&
  {Jenness}}]{Hughes:1998p9666}
{Hughes}, D.~H., {Serjeant}, S., {Dunlop}, J., {et~al.} 1998,
  \href{http://dx.doi.org/10.1038/28328}{\JournalTitle{\nat}, 394, 241}

\bibitem[{{Hung} {et~al.}(2015){Hung}, {Rich}, {Yuan}, {Larson}, {Casey},
  {Smith}, {Sanders}, {Kewley}, \& {Hayward}}]{Hung:2015aa}
{Hung}, C.-L., {Rich}, J.~A., {Yuan}, T., {et~al.} 2015,
  \href{http://dx.doi.org/10.1088/0004-637X/803/2/62}{\JournalTitle{\apj}, 803,
  62}

\bibitem[{Hurvich \& Tsai(1989)}]{Hurvich:1989aa}
Hurvich, C.~M., \& Tsai, C.-L. 1989,
  \href{http://dx.doi.org/10.1093/biomet/76.2.297}{\JournalTitle{Biometrika},
  76, 297}

\bibitem[{{Huynh} {et~al.}(2017){Huynh}, {Emonts}, {Kimball}, {Seymour},
  {Smail}, {Swinbank}, {Brandt}, {Casey}, {Chapman}, {Dannerbauer}, {Hodge},
  {Ivison}, {Schinnerer}, {Thomson}, {van der Werf}, \&
  {Wardlow}}]{Huynh:2017aa}
{Huynh}, M.~T., {Emonts}, B.~H.~C., {Kimball}, A.~E., {et~al.} 2017,
  \JournalTitle{MNRAS in press},
  \href{http://arxiv.org/abs/1701.05698}{{\sffamily arXiv:1701.05698}}

\bibitem[{{Ikarashi} {et~al.}(2015){Ikarashi}, {Ivison}, {Caputi}, {Aretxaga},
  {Dunlop}, {Hatsukade}, {Hughes}, {Iono}, {Izumi}, {Kawabe}, {Kohno}, {Lagos},
  {Motohara}, {Nakanishi}, {Ohta}, {Tamura}, {Umehata}, {Wilson}, {Yabe}, \&
  {Yun}}]{Ikarashi:2015aa}
{Ikarashi}, S., {Ivison}, R.~J., {Caputi}, K.~I., {et~al.} 2015,
  \href{http://dx.doi.org/10.1088/0004-637X/810/2/133}{\JournalTitle{\apj},
  810, 133}

\bibitem[{{Iono} {et~al.}(2007){Iono}, {Wilson}, {Takakuwa}, {Yun}, {Petitpas},
  {Peck}, {Ho}, {Matsushita}, {Pihlstrom}, \& {Wang}}]{Iono:2007aa}
{Iono}, D., {Wilson}, C.~D., {Takakuwa}, S., {et~al.} 2007,
  \href{http://dx.doi.org/10.1086/512362}{\JournalTitle{\apj}, 659, 283}

\bibitem[{{Ivison} {et~al.}(2011){Ivison}, {Papadopoulos}, {Smail}, {Greve},
  {Thomson}, {Xilouris}, \& {Chapman}}]{Ivison:2011aa}
{Ivison}, R.~J., {Papadopoulos}, P.~P., {Smail}, I., {et~al.} 2011,
  \href{http://dx.doi.org/10.1111/j.1365-2966.2010.18028.x}{\JournalTitle{\mnras},
  412, 1913}

\bibitem[{{Karim} {et~al.}(2013){Karim}, {Swinbank}, {Hodge}, {Smail},
  {Walter}, {Biggs}, {Simpson}, {Danielson}, {Alexander}, {Bertoldi}, {de
  Breuck}, {Chapman}, {Coppin}, {Dannerbauer}, {Edge}, {Greve}, {Ivison},
  {Knudsen}, {Menten}, {Schinnerer}, {Wardlow}, {Wei{\ss}}, \& {van der
  Werf}}]{Karim:2013fk}
{Karim}, A., {Swinbank}, A.~M., {Hodge}, J.~A., {et~al.} 2013,
  \href{http://dx.doi.org/10.1093/mnras/stt196}{\JournalTitle{\mnras}, 432, 2}

\bibitem[{{Kennicutt} \& {Evans}(2012)}]{Kennicutt:2012aa}
{Kennicutt}, R.~C., \& {Evans}, N.~J. 2012,
  \href{http://dx.doi.org/10.1146/annurev-astro-081811-125610}{\JournalTitle{\araa},
  50, 531}

\bibitem[{{Kennicutt} {et~al.}(2011){Kennicutt}, {Calzetti}, {Aniano},
  {Appleton}, {Armus}, {Beir{\~a}o}, {Bolatto}, {Brandl}, {Crocker}, {Croxall},
  {Dale}, {Donovan Meyer}, {Draine}, {Engelbracht}, {Galametz}, {Gordon},
  {Groves}, {Hao}, {Helou}, {Hinz}, {Hunt}, {Johnson}, {Koda}, {Krause},
  {Leroy}, {Li}, {Meidt}, {Montiel}, {Murphy}, {Rahman}, {Rix}, {Roussel},
  {Sandstrom}, {Sauvage}, {Schinnerer}, {Skibba}, {Smith}, {Srinivasan},
  {Vigroux}, {Walter}, {Wilson}, {Wolfire}, \& {Zibetti}}]{Kennicutt:2011aa}
{Kennicutt}, R.~C., {Calzetti}, D., {Aniano}, G., {et~al.} 2011,
  \href{http://dx.doi.org/10.1086/663818}{\JournalTitle{\pasp}, 123, 1347}

\bibitem[{{Kennicutt}(1989)}]{Kennicutt:1989aa}
{Kennicutt}, Jr., R.~C. 1989,
  \href{http://dx.doi.org/10.1086/167834}{\JournalTitle{\apj}, 344, 685}

\bibitem[{{Kennicutt}(1998{\natexlab{a}})}]{Kennicutt:1998aa}
---. 1998{\natexlab{a}},
  \href{http://dx.doi.org/10.1146/annurev.astro.36.1.189}{\JournalTitle{\araa},
  36, 189}

\bibitem[{{Kennicutt}(1998{\natexlab{b}})}]{Kennicutt:1998p5718}
---. 1998{\natexlab{b}},
  \href{http://dx.doi.org/10.1086/305588}{\JournalTitle{\apj}, 498, 541}

\bibitem[{{Kennicutt} {et~al.}(2003){Kennicutt}, {Armus}, {Bendo}, {Calzetti},
  {Dale}, {Draine}, {Engelbracht}, {Gordon}, {Grauer}, {Helou}, {Hollenbach},
  {Jarrett}, {Kewley}, {Leitherer}, {Li}, {Malhotra}, {Regan}, {Rieke},
  {Rieke}, {Roussel}, {Smith}, {Thornley}, \& {Walter}}]{Kennicutt:2003aa}
{Kennicutt}, Jr., R.~C., {Armus}, L., {Bendo}, G., {et~al.} 2003,
  \href{http://dx.doi.org/10.1086/376941}{\JournalTitle{\pasp}, 115, 928}

\bibitem[{{Koekemoer} {et~al.}(2011){Koekemoer}, {Faber}, {Ferguson}, {Grogin},
  {Kocevski}, {Koo}, {Lai}, {Lotz}, {Lucas}, {McGrath}, {Ogaz}, {Rajan},
  {Riess}, {Rodney}, {Strolger}, {Casertano}, {Castellano}, {Dahlen},
  {Dickinson}, {Dolch}, {Fontana}, {Giavalisco}, {Grazian}, {Guo}, {Hathi},
  {Huang}, {van der Wel}, {Yan}, {Acquaviva}, {Alexander}, {Almaini}, {Ashby},
  {Barden}, {Bell}, {Bournaud}, {Brown}, {Caputi}, {Cassata}, {Challis},
  {Chary}, {Cheung}, {Cirasuolo}, {Conselice}, {Roshan Cooray}, {Croton},
  {Daddi}, {Dav{\'e}}, {de Mello}, {de Ravel}, {Dekel}, {Donley}, {Dunlop},
  {Dutton}, {Elbaz}, {Fazio}, {Filippenko}, {Finkelstein}, {Frazer}, {Gardner},
  {Garnavich}, {Gawiser}, {Gruetzbauch}, {Hartley}, {H{\"a}ussler},
  {Herrington}, {Hopkins}, {Huang}, {Jha}, {Johnson}, {Kartaltepe},
  {Khostovan}, {Kirshner}, {Lani}, {Lee}, {Li}, {Madau}, {McCarthy},
  {McIntosh}, {McLure}, {McPartland}, {Mobasher}, {Moreira}, {Mortlock},
  {Moustakas}, {Mozena}, {Nandra}, {Newman}, {Nielsen}, {Niemi}, {Noeske},
  {Papovich}, {Pentericci}, {Pope}, {Primack}, {Ravindranath}, {Reddy},
  {Renzini}, {Rix}, {Robaina}, {Rosario}, {Rosati}, {Salimbeni}, {Scarlata},
  {Siana}, {Simard}, {Smidt}, {Snyder}, {Somerville}, {Spinrad}, {Straughn},
  {Telford}, {Teplitz}, {Trump}, {Vargas}, {Villforth}, {Wagner}, {Wandro},
  {Wechsler}, {Weiner}, {Wiklind}, {Wild}, {Wilson}, {Wuyts}, \&
  {Yun}}]{Koekemoer:2011aa}
{Koekemoer}, A.~M., {Faber}, S.~M., {Ferguson}, H.~C., {et~al.} 2011,
  \href{http://dx.doi.org/10.1088/0067-0049/197/2/36}{\JournalTitle{\apjs},
  197, 36}

\bibitem[{{Kong} {et~al.}(2004){Kong}, {Charlot}, {Brinchmann}, \&
  {Fall}}]{Kong:2004aa}
{Kong}, X., {Charlot}, S., {Brinchmann}, J., \& {Fall}, S.~M. 2004,
  \href{http://dx.doi.org/10.1111/j.1365-2966.2004.07556.x}{\JournalTitle{\mnras},
  349, 769}

\bibitem[{{Koprowski} {et~al.}(2016){Koprowski}, {Coppin}, {Geach}, {Hine},
  {Bremer}, {Chapman}, {Davies}, {Hayashino}, {Knudsen}, {Kubo}, {Lehmer},
  {Matsuda}, {Smith}, {van der Werf}, {Violino}, \&
  {Yamada}}]{Koprowski:2016ab}
{Koprowski}, M.~P., {Coppin}, K.~E.~K., {Geach}, J.~E., {et~al.} 2016,
  \href{http://dx.doi.org/10.3847/2041-8205/828/2/L21}{\JournalTitle{\apjl},
  828, L21}

\bibitem[{{Krajnovi{\'c}} {et~al.}(2006){Krajnovi{\'c}}, {Cappellari}, {de
  Zeeuw}, \& {Copin}}]{Krajnovic:2006aa}
{Krajnovi{\'c}}, D., {Cappellari}, M., {de Zeeuw}, P.~T., \& {Copin}, Y. 2006,
  \href{http://dx.doi.org/10.1111/j.1365-2966.2005.09902.x}{\JournalTitle{\mnras},
  366, 787}

\bibitem[{{Leroy} {et~al.}(2008){Leroy}, {Walter}, {Brinks}, {Bigiel}, {de
  Blok}, {Madore}, \& {Thornley}}]{Leroy:2008aa}
{Leroy}, A.~K., {Walter}, F., {Brinks}, E., {et~al.} 2008,
  \href{http://dx.doi.org/10.1088/0004-6256/136/6/2782}{\JournalTitle{\aj},
  136, 2782}

\bibitem[{{Leroy} {et~al.}(2009){Leroy}, {Walter}, {Bigiel}, {Usero}, {Weiss},
  {Brinks}, {de Blok}, {Kennicutt}, {Schuster}, {Kramer}, {Wiesemeyer}, \&
  {Roussel}}]{Leroy:2009aa}
{Leroy}, A.~K., {Walter}, F., {Bigiel}, F., {et~al.} 2009,
  \href{http://dx.doi.org/10.1088/0004-6256/137/6/4670}{\JournalTitle{\aj},
  137, 4670}

\bibitem[{{Leroy} {et~al.}(2011){Leroy}, {Bolatto}, {Gordon}, {Sandstrom},
  {Gratier}, {Rosolowsky}, {Engelbracht}, {Mizuno}, {Corbelli}, {Fukui}, \&
  {Kawamura}}]{Leroy:2011aa}
{Leroy}, A.~K., {Bolatto}, A., {Gordon}, K., {et~al.} 2011,
  \href{http://dx.doi.org/10.1088/0004-637X/737/1/12}{\JournalTitle{\apj}, 737,
  12}

\bibitem[{{Madau} \& {Dickinson}(2014)}]{Madau:2014aa}
{Madau}, P., \& {Dickinson}, M. 2014,
  \href{http://dx.doi.org/10.1146/annurev-astro-081811-125615}{\JournalTitle{\araa},
  52, 415}

\bibitem[{{Men{\'e}ndez-Delmestre} {et~al.}(2013){Men{\'e}ndez-Delmestre},
  {Blain}, {Swinbank}, {Smail}, {Ivison}, {Chapman}, \& {Gon{\c
  c}alves}}]{Menendez-Delmestre:2013aa}
{Men{\'e}ndez-Delmestre}, K., {Blain}, A.~W., {Swinbank}, M., {et~al.} 2013,
  \href{http://dx.doi.org/10.1088/0004-637X/767/2/151}{\JournalTitle{\apj},
  767, 151}

\bibitem[{{Meurer} {et~al.}(1999){Meurer}, {Heckman}, \&
  {Calzetti}}]{Meurer:1999aa}
{Meurer}, G.~R., {Heckman}, T.~M., \& {Calzetti}, D. 1999,
  \href{http://dx.doi.org/10.1086/307523}{\JournalTitle{\apj}, 521, 64}

\bibitem[{{Miller} {et~al.}(2011){Miller}, {Bundy}, {Sullivan}, {Ellis}, \&
  {Treu}}]{Miller:2011aa}
{Miller}, S.~H., {Bundy}, K., {Sullivan}, M., {Ellis}, R.~S., \& {Treu}, T.
  2011,
  \href{http://dx.doi.org/10.1088/0004-637X/741/2/115}{\JournalTitle{\apj},
  741, 115}

\bibitem[{{Molina} {et~al.}(2017){Molina}, {Ibar}, {Swinbank}, {Sobral},
  {Best}, {Smail}, {Escala}, \& {Cirasuolo}}]{Molina:2017aa}
{Molina}, J., {Ibar}, E., {Swinbank}, A.~M., {et~al.} 2017,
  \href{http://dx.doi.org/10.1093/mnras/stw3120}{\JournalTitle{\mnras}, 466,
  892}

\bibitem[{{Narayanan} {et~al.}(2017){Narayanan}, {Dave}, {Johnson}, {Thompson},
  {Conroy}, \& {Geach}}]{Narayanan:2017aa}
{Narayanan}, D., {Dave}, R., {Johnson}, B., {et~al.} 2017, \JournalTitle{MNRAS
  submitted}, \href{http://arxiv.org/abs/1705.05858}{{\sffamily
  arXiv:1705.05858}}

\bibitem[{{Narayanan} {et~al.}(2010){Narayanan}, {Hayward}, {Cox}, {Hernquist},
  {Jonsson}, {Younger}, \& {Groves}}]{Narayanan:2010aa}
{Narayanan}, D., {Hayward}, C.~C., {Cox}, T.~J., {et~al.} 2010,
  \JournalTitle{\mnras}, 401, 1613

\bibitem[{{Olivares} {et~al.}(2016){Olivares}, {Treister}, {Privon},
  {Alaghband-Zadeh}, {Casey}, {Schawinski}, {Kurczynski}, {Gawiser}, {Nagar},
  {Chapman}, {Bauer}, \& {Sanders}}]{Olivares:2016aa}
{Olivares}, V., {Treister}, E., {Privon}, G.~C., {et~al.} 2016,
  \href{http://dx.doi.org/10.3847/0004-637X/827/1/57}{\JournalTitle{\apj}, 827,
  57}

\bibitem[{{Osterbrock}(1989)}]{1989agna.book.....O}
{Osterbrock}, D.~E. 1989, {Astrophysics of gaseous nebulae and active galactic
  nuclei}

\bibitem[{{Overzier} {et~al.}(2011){Overzier}, {Heckman}, {Wang}, {Armus},
  {Buat}, {Howell}, {Meurer}, {Seibert}, {Siana}, {Basu-Zych}, {Charlot},
  {Gon{\c c}alves}, {Martin}, {Neill}, {Rich}, {Salim}, \&
  {Schiminovich}}]{Overzier:2011aa}
{Overzier}, R.~A., {Heckman}, T.~M., {Wang}, J., {et~al.} 2011,
  \href{http://dx.doi.org/10.1088/2041-8205/726/1/L7}{\JournalTitle{\apjl},
  726, L7}

\bibitem[{{Planck Collaboration} {et~al.}(2014){Planck Collaboration}, {Ade},
  {Aghanim}, {Armitage-Caplan}, {Arnaud}, {Ashdown}, {Atrio-Barandela},
  {Aumont}, {Baccigalupi}, {Banday}, \& et~al.}]{Planck-Collaboration:2014aa}
{Planck Collaboration}, {Ade}, P.~A.~R., {Aghanim}, N., {et~al.} 2014,
  \href{http://dx.doi.org/10.1051/0004-6361/201321591}{\JournalTitle{\aap},
  571, A16}

\bibitem[{{Popping} {et~al.}(2017){Popping}, {Puglisi}, \&
  {Norman}}]{Popping:2017aa}
{Popping}, G., {Puglisi}, A., \& {Norman}, C.~A. 2017, \JournalTitle{MNRAS
  submitted}, \href{http://arxiv.org/abs/1706.06587}{{\sffamily
  arXiv:1706.06587}}

\bibitem[{{Rawle} {et~al.}(2014){Rawle}, {Egami}, {Bussmann}, {Gurwell},
  {Ivison}, {Boone}, {Combes}, {Danielson}, {Rex}, {Richard}, {Smail},
  {Swinbank}, {Altieri}, {Blain}, {Clement}, {Dessauges-Zavadsky}, {Edge},
  {Fazio}, {Jones}, {Kneib}, {Omont}, {P{\'e}rez-Gonz{\'a}lez}, {Schaerer},
  {Valtchanov}, {van der Werf}, {Walth}, {Zamojski}, \&
  {Zemcov}}]{Rawle:2014aa}
{Rawle}, T.~D., {Egami}, E., {Bussmann}, R.~S., {et~al.} 2014,
  \href{http://dx.doi.org/10.1088/0004-637X/783/1/59}{\JournalTitle{\apj}, 783,
  59}

\bibitem[{{Reddy} {et~al.}(2012){Reddy}, {Dickinson}, {Elbaz}, {Morrison},
  {Giavalisco}, {Ivison}, {Papovich}, {Scott}, {Buat}, {Burgarella},
  {Charmandaris}, {Daddi}, {Magdis}, {Murphy}, {Altieri}, {Aussel},
  {Dannerbauer}, {Dasyra}, {Hwang}, {Kartaltepe}, {Leiton}, {Magnelli}, \&
  {Popesso}}]{Reddy:2012aa}
{Reddy}, N., {Dickinson}, M., {Elbaz}, D., {et~al.} 2012,
  \href{http://dx.doi.org/10.1088/0004-637X/744/2/154}{\JournalTitle{\apj},
  744, 154}

\bibitem[{{Riechers} {et~al.}(2011){Riechers}, {Hodge}, {Walter}, {Carilli}, \&
  {Bertoldi}}]{Riechers:2011aa}
{Riechers}, D.~A., {Hodge}, J., {Walter}, F., {Carilli}, C.~L., \& {Bertoldi},
  F. 2011,
  \href{http://dx.doi.org/10.1088/2041-8205/739/1/L31}{\JournalTitle{\apjl},
  739, L31}

\bibitem[{{Rousselot} {et~al.}(2000){Rousselot}, {Lidman}, {Cuby}, {Moreels},
  \& {Monnet}}]{Rousselot:2000aa}
{Rousselot}, P., {Lidman}, C., {Cuby}, J.-G., {Moreels}, G., \& {Monnet}, G.
  2000, \JournalTitle{\aap}, 354, 1134

\bibitem[{{Sakamoto} {et~al.}(2006){Sakamoto}, {Ho}, \&
  {Peck}}]{Sakamoto:2006aa}
{Sakamoto}, K., {Ho}, P.~T.~P., \& {Peck}, A.~B. 2006,
  \href{http://dx.doi.org/10.1086/503827}{\JournalTitle{\apj}, 644, 862}

\bibitem[{{Sanders} \& {Mirabel}(1996)}]{Sanders:1996p6419}
{Sanders}, D.~B., \& {Mirabel}, I.~F. 1996,
  \href{http://dx.doi.org/10.1146/annurev.astro.34.1.749}{\JournalTitle{\araa},
  34, 749}

\bibitem[{{Sandstrom} {et~al.}(2013){Sandstrom}, {Leroy}, {Walter}, {Bolatto},
  {Croxall}, {Draine}, {Wilson}, {Wolfire}, {Calzetti}, {Kennicutt}, {Aniano},
  {Donovan Meyer}, {Usero}, {Bigiel}, {Brinks}, {de Blok}, {Crocker}, {Dale},
  {Engelbracht}, {Galametz}, {Groves}, {Hunt}, {Koda}, {Kreckel}, {Linz},
  {Meidt}, {Pellegrini}, {Rix}, {Roussel}, {Schinnerer}, {Schruba}, {Schuster},
  {Skibba}, {van der Laan}, {Appleton}, {Armus}, {Brandl}, {Gordon}, {Hinz},
  {Krause}, {Montiel}, {Sauvage}, {Schmiedeke}, {Smith}, \&
  {Vigroux}}]{Sandstrom:2013aa}
{Sandstrom}, K.~M., {Leroy}, A.~K., {Walter}, F., {et~al.} 2013,
  \href{http://dx.doi.org/10.1088/0004-637X/777/1/5}{\JournalTitle{\apj}, 777,
  5}

\bibitem[{{Schmidt}(1959)}]{Schmidt:1959aa}
{Schmidt}, M. 1959,
  \href{http://dx.doi.org/10.1086/146614}{\JournalTitle{\apj}, 129, 243}

\bibitem[{{Scoville} {et~al.}(1991){Scoville}, {Sargent}, {Sanders}, \&
  {Soifer}}]{Scoville:1991aa}
{Scoville}, N.~Z., {Sargent}, A.~I., {Sanders}, D.~B., \& {Soifer}, B.~T. 1991,
  \href{http://dx.doi.org/10.1086/185897}{\JournalTitle{\apjl}, 366, L5}

\bibitem[{{Scoville} {et~al.}(1997){Scoville}, {Yun}, \&
  {Bryant}}]{Scoville:1997aa}
{Scoville}, N.~Z., {Yun}, M.~S., \& {Bryant}, P.~M. 1997,
  \href{http://dx.doi.org/10.1086/304368}{\JournalTitle{\apj}, 484, 702}

\bibitem[{{Shapiro} {et~al.}(2008){Shapiro}, {Genzel}, {F{\"o}rster Schreiber},
  {Tacconi}, {Bouch{\'e}}, {Cresci}, {Davies}, {Eisenhauer}, {Johansson},
  {Krajnovi{\'c}}, {Lutz}, {Naab}, {Arimoto}, {Arribas}, {Cimatti}, {Colina},
  {Daddi}, {Daigle}, {Erb}, {Hernandez}, {Kong}, {Mignoli}, {Onodera},
  {Renzini}, {Shapley}, \& {Steidel}}]{Shapiro:2008aa}
{Shapiro}, K.~L., {Genzel}, R., {F{\"o}rster Schreiber}, N.~M., {et~al.} 2008,
  \href{http://dx.doi.org/10.1086/587133}{\JournalTitle{\apj}, 682, 231}

\bibitem[{{Sharon} {et~al.}(2013){Sharon}, {Baker}, {Harris}, \&
  {Thomson}}]{Sharon:2013aa}
{Sharon}, C.~E., {Baker}, A.~J., {Harris}, A.~I., \& {Thomson}, A.~P. 2013,
  \href{http://dx.doi.org/10.1088/0004-637X/765/1/6}{\JournalTitle{\apj}, 765,
  6}

\bibitem[{{Sharon} {et~al.}(2016){Sharon}, {Riechers}, {Hodge}, {Carilli},
  {Walter}, {Wei{\ss}}, {Knudsen}, \& {Wagg}}]{Sharon:2016aa}
{Sharon}, C.~E., {Riechers}, D.~A., {Hodge}, J., {et~al.} 2016,
  \href{http://dx.doi.org/10.3847/0004-637X/827/1/18}{\JournalTitle{\apj}, 827,
  18}

\bibitem[{{Simpson} {et~al.}(2014){Simpson}, {Swinbank}, {Smail}, {Alexander},
  {Brandt}, {Bertoldi}, {de Breuck}, {Chapman}, {Coppin}, {da Cunha},
  {Danielson}, {Dannerbauer}, {Greve}, {Hodge}, {Ivison}, {Karim}, {Knudsen},
  {Poggianti}, {Schinnerer}, {Thomson}, {Walter}, {Wardlow}, {Wei{\ss}}, \&
  {van der Werf}}]{Simpson:2014aa}
{Simpson}, J.~M., {Swinbank}, A.~M., {Smail}, I., {et~al.} 2014,
  \href{http://dx.doi.org/10.1088/0004-637X/788/2/125}{\JournalTitle{\apj},
  788, 125}

\bibitem[{{Simpson} {et~al.}(2015){Simpson}, {Smail}, {Swinbank}, {Almaini},
  {Blain}, {Bremer}, {Chapman}, {Chen}, {Conselice}, {Coppin}, {Danielson},
  {Dunlop}, {Edge}, {Farrah}, {Geach}, {Hartley}, {Ivison}, {Karim}, {Lani},
  {Ma}, {Meijerink}, {Micha{\l}owski}, {Mortlock}, {Scott}, {Simpson},
  {Spaans}, {Thomson}, {van Kampen}, \& {van der Werf}}]{Simpson:2015aa}
{Simpson}, J.~M., {Smail}, I., {Swinbank}, A.~M., {et~al.} 2015,
  \href{http://dx.doi.org/10.1088/0004-637X/799/1/81}{\JournalTitle{\apj}, 799,
  81}

\bibitem[{{Simpson} {et~al.}(2017){Simpson}, {Smail}, {Swinbank}, {Ivison},
  {Dunlop}, {Geach}, {Almaini}, {Arumugam}, {Bremer}, {Chen}, {Conselice},
  {Coppin}, {Farrah}, {Ibar}, {Hartley}, {Ma}, {Micha{\l}owski}, {Scott},
  {Spaans}, {Thomson}, \& {van der Werf}}]{Simpson:2017ab}
---. 2017,
  \href{http://dx.doi.org/10.3847/1538-4357/aa65d0}{\JournalTitle{\apj}, 839,
  58}

\bibitem[{{Siringo} {et~al.}(2009){Siringo}, {Kreysa}, {Kov{\'a}cs},
  {Schuller}, {Wei{\ss}}, {Esch}, {Gem{\"u}nd}, {Jethava}, {Lundershausen},
  {Colin}, {G{\"u}sten}, {Menten}, {Beelen}, {Bertoldi}, {Beeman}, \&
  {Haller}}]{Siringo:2009rt}
{Siringo}, G., {Kreysa}, E., {Kov{\'a}cs}, A., {et~al.} 2009,
  \href{http://dx.doi.org/10.1051/0004-6361/200811454}{\JournalTitle{\aap},
  497, 945}

\bibitem[{{Skelton} {et~al.}(2014){Skelton}, {Whitaker}, {Momcheva}, {Brammer},
  {van Dokkum}, {Labb{\'e}}, {Franx}, {van der Wel}, {Bezanson}, {Da Cunha},
  {Fumagalli}, {F{\"o}rster Schreiber}, {Kriek}, {Leja}, {Lundgren}, {Magee},
  {Marchesini}, {Maseda}, {Nelson}, {Oesch}, {Pacifici}, {Patel}, {Price},
  {Rix}, {Tal}, {Wake}, \& {Wuyts}}]{Skelton:2014aa}
{Skelton}, R.~E., {Whitaker}, K.~E., {Momcheva}, I.~G., {et~al.} 2014,
  \href{http://dx.doi.org/10.1088/0067-0049/214/2/24}{\JournalTitle{\apjs},
  214, 24}

\bibitem[{{Smail} {et~al.}(1997){Smail}, {Ivison}, \&
  {Blain}}]{Smail:1997p6820}
{Smail}, I., {Ivison}, R.~J., \& {Blain}, A.~W. 1997,
  \href{http://dx.doi.org/10.1086/311017}{\JournalTitle{\apjl}, 490, L5}

\bibitem[{{Spilker} {et~al.}(2014){Spilker}, {Marrone}, {Aguirre}, {Aravena},
  {Ashby}, {B{\'e}thermin}, {Bradford}, {Bothwell}, {Brodwin}, {Carlstrom},
  {Chapman}, {Crawford}, {de Breuck}, {Fassnacht}, {Gonzalez}, {Greve},
  {Gullberg}, {Hezaveh}, {Holzapfel}, {Husband}, {Ma}, {Malkan}, {Murphy},
  {Reichardt}, {Rotermund}, {Stalder}, {Stark}, {Strandet}, {Vieira},
  {Wei{\ss}}, \& {Welikala}}]{Spilker:2014aa}
{Spilker}, J.~S., {Marrone}, D.~P., {Aguirre}, J.~E., {et~al.} 2014,
  \href{http://dx.doi.org/10.1088/0004-637X/785/2/149}{\JournalTitle{\apj},
  785, 149}

\bibitem[{{Spilker} {et~al.}(2015){Spilker}, {Aravena}, {Marrone},
  {B{\'e}thermin}, {Bothwell}, {Carlstrom}, {Chapman}, {Collier}, {de Breuck},
  {Fassnacht}, {Galvin}, {Gonzalez}, {Gonz{\'a}lez-L{\'o}pez}, {Grieve},
  {Hezaveh}, {Ma}, {Malkan}, {O'Brien}, {Rotermund}, {Strandet}, {Vieira},
  {Weiss}, \& {Wong}}]{Spilker:2015aa}
{Spilker}, J.~S., {Aravena}, M., {Marrone}, D.~P., {et~al.} 2015,
  \href{http://dx.doi.org/10.1088/0004-637X/811/2/124}{\JournalTitle{\apj},
  811, 124}

\bibitem[{{Spilker} {et~al.}(2016){Spilker}, {Marrone}, {Aravena},
  {B{\'e}thermin}, {Bothwell}, {Carlstrom}, {Chapman}, {Crawford}, {de Breuck},
  {Fassnacht}, {Gonzalez}, {Greve}, {Hezaveh}, {Litke}, {Ma}, {Malkan},
  {Rotermund}, {Strandet}, {Vieira}, {Weiss}, \& {Welikala}}]{Spilker:2016aa}
{Spilker}, J.~S., {Marrone}, D.~P., {Aravena}, M., {et~al.} 2016,
  \href{http://dx.doi.org/10.3847/0004-637X/826/2/112}{\JournalTitle{\apj},
  826, 112}

\bibitem[{{Swinbank} {et~al.}(2006){Swinbank}, {Chapman}, {Smail}, {Lindner},
  {Borys}, {Blain}, {Ivison}, \& {Lewis}}]{Swinbank:2006aa}
{Swinbank}, A.~M., {Chapman}, S.~C., {Smail}, I., {et~al.} 2006,
  \href{http://dx.doi.org/10.1111/j.1365-2966.2006.10673.x}{\JournalTitle{\mnras},
  371, 465}

\bibitem[{{Swinbank} {et~al.}(2012){Swinbank}, {Sobral}, {Smail}, {Geach},
  {Best}, {McCarthy}, {Crain}, \& {Theuns}}]{Swinbank:2012aa}
{Swinbank}, A.~M., {Sobral}, D., {Smail}, I., {et~al.} 2012,
  \href{http://dx.doi.org/10.1111/j.1365-2966.2012.21774.x}{\JournalTitle{\mnras},
  426, 935}

\bibitem[{{Swinbank} {et~al.}(2010){Swinbank}, {Smail}, {Chapman}, {Borys},
  {Alexander}, {Blain}, {Conselice}, {Hainline}, \& {Ivison}}]{Swinbank:2010aa}
{Swinbank}, A.~M., {Smail}, I., {Chapman}, S.~C., {et~al.} 2010,
  \href{http://dx.doi.org/10.1111/j.1365-2966.2010.16485.x}{\JournalTitle{\mnras},
  405, 234}

\bibitem[{{Swinbank} {et~al.}(2014){Swinbank}, {Simpson}, {Smail}, {Harrison},
  {Hodge}, {Karim}, {Walter}, {Alexander}, {Brandt}, {de Breuck}, {da Cunha},
  {Chapman}, {Coppin}, {Danielson}, {Dannerbauer}, {Decarli}, {Greve},
  {Ivison}, {Knudsen}, {Lagos}, {Schinnerer}, {Thomson}, {Wardlow}, {Wei{\ss}},
  \& {van der Werf}}]{Swinbank:2014aa}
{Swinbank}, A.~M., {Simpson}, J.~M., {Smail}, I., {et~al.} 2014,
  \href{http://dx.doi.org/10.1093/mnras/stt2273}{\JournalTitle{\mnras}, 438,
  1267}

\bibitem[{{Tacconi} {et~al.}(2013){Tacconi}, {Neri}, {Genzel}, {Combes},
  {Bolatto}, {Cooper}, {Wuyts}, {Bournaud}, {Burkert}, {Comerford}, {Cox},
  {Davis}, {F{\"o}rster Schreiber}, {Garc{\'{\i}}a-Burillo}, {Gracia-Carpio},
  {Lutz}, {Naab}, {Newman}, {Omont}, {Saintonge}, {Shapiro Griffin}, {Shapley},
  {Sternberg}, \& {Weiner}}]{Tacconi:2013aa}
{Tacconi}, L.~J., {Neri}, R., {Genzel}, R., {et~al.} 2013,
  \href{http://dx.doi.org/10.1088/0004-637X/768/1/74}{\JournalTitle{\apj}, 768,
  74}

\bibitem[{{Tadaki} {et~al.}(2017){Tadaki}, {Kodama}, {Nelson}, {Belli},
  {F{\"o}rster Schreiber}, {Genzel}, {Hayashi}, {Herrera-Camus}, {Koyama},
  {Lang}, {Lutz}, {Shimakawa}, {Tacconi}, {{\"U}bler}, {Wisnioski}, {Wuyts},
  {Hatsukade}, {Lippa}, {Nakanishi}, {Ikarashi}, {Kohno}, {Suzuki}, {Tamura},
  \& {Tanaka}}]{Tadaki:2017ab}
{Tadaki}, K.-i., {Kodama}, T., {Nelson}, E.~J., {et~al.} 2017,
  \href{http://dx.doi.org/10.3847/2041-8213/aa7338}{\JournalTitle{\apjl}, 841,
  L25}

\bibitem[{{Takeuchi} {et~al.}(2012){Takeuchi}, {Yuan}, {Ikeyama}, {Murata}, \&
  {Inoue}}]{Takeuchi:2012aa}
{Takeuchi}, T.~T., {Yuan}, F.-T., {Ikeyama}, A., {Murata}, K.~L., \& {Inoue},
  A.~K. 2012,
  \href{http://dx.doi.org/10.1088/0004-637X/755/2/144}{\JournalTitle{\apj},
  755, 144}

\bibitem[{{Thomson} {et~al.}(2015){Thomson}, {Ivison}, {Owen}, {Danielson},
  {Swinbank}, \& {Smail}}]{Thomson:2015aa}
{Thomson}, A.~P., {Ivison}, R.~J., {Owen}, F.~N., {et~al.} 2015,
  \href{http://dx.doi.org/10.1093/mnras/stv118}{\JournalTitle{\mnras}, 448,
  1874}

\bibitem[{{To} {et~al.}(2014){To}, {Wang}, \& {Owen}}]{To:2014aa}
{To}, C.-H., {Wang}, W.-H., \& {Owen}, F.~N. 2014,
  \href{http://dx.doi.org/10.1088/0004-637X/792/2/139}{\JournalTitle{\apj},
  792, 139}

\bibitem[{{Ueda} {et~al.}(2014){Ueda}, {Iono}, {Yun}, {Crocker}, {Narayanan},
  {Komugi}, {Espada}, {Hatsukade}, {Kaneko}, {Matsuda}, {Tamura}, {Wilner},
  {Kawabe}, \& {Pan}}]{Ueda:2014aa}
{Ueda}, J., {Iono}, D., {Yun}, M.~S., {et~al.} 2014,
  \href{http://dx.doi.org/10.1088/0067-0049/214/1/1}{\JournalTitle{\apjs}, 214,
  1}

\bibitem[{{van der Wel} {et~al.}(2014){van der Wel}, {Franx}, {van Dokkum},
  {Skelton}, {Momcheva}, {Whitaker}, {Brammer}, {Bell}, {Rix}, {Wuyts},
  {Ferguson}, {Holden}, {Barro}, {Koekemoer}, {Chang}, {McGrath},
  {H{\"a}ussler}, {Dekel}, {Behroozi}, {Fumagalli}, {Leja}, {Lundgren},
  {Maseda}, {Nelson}, {Wake}, {Patel}, {Labb{\'e}}, {Faber}, {Grogin}, \&
  {Kocevski}}]{van-der-Wel:2014aa}
{van der Wel}, A., {Franx}, M., {van Dokkum}, P.~G., {et~al.} 2014,
  \href{http://dx.doi.org/10.1088/0004-637X/788/1/28}{\JournalTitle{\apj}, 788,
  28}

\bibitem[{{Walter} {et~al.}(2008){Walter}, {Brinks}, {de Blok}, {Bigiel},
  {Kennicutt}, {Thornley}, \& {Leroy}}]{Walter:2008aa}
{Walter}, F., {Brinks}, E., {de Blok}, W.~J.~G., {et~al.} 2008,
  \href{http://dx.doi.org/10.1088/0004-6256/136/6/2563}{\JournalTitle{\aj},
  136, 2563}

\bibitem[{{Wang} {et~al.}(2013){Wang}, {Brandt}, {Luo}, {Smail}, {Alexander},
  {Danielson}, {Hodge}, {Karim}, {Lehmer}, {Simpson}, {Swinbank}, {Walter},
  {Wardlow}, {Xue}, {Chapman}, {Coppin}, {Dannerbauer}, {De Breuck}, {Menten},
  \& {van der Werf}}]{Wang:2013aa}
{Wang}, S.~X., {Brandt}, W.~N., {Luo}, B., {et~al.} 2013,
  \href{http://dx.doi.org/10.1088/0004-637X/778/2/179}{\JournalTitle{\apj},
  778, 179}

\bibitem[{{Wardlow} {et~al.}(2011){Wardlow}, {Smail}, {Coppin}, {Alexander},
  {Brandt}, {Danielson}, {Luo}, {Swinbank}, {Walter}, {Wei{\ss}}, {Xue},
  {Zibetti}, {Bertoldi}, {Biggs}, {Chapman}, {Dannerbauer}, {Dunlop},
  {Gawiser}, {Ivison}, {Knudsen}, {Kov{\'a}cs}, {Lacey}, {Menten}, {Padilla},
  {Rix}, \& {van der Werf}}]{Wardlow:2011qy}
{Wardlow}, J.~L., {Smail}, I., {Coppin}, K.~E.~K., {et~al.} 2011,
  \href{http://dx.doi.org/10.1111/j.1365-2966.2011.18795.x}{\JournalTitle{\mnras},
  415, 1479}

\bibitem[{{Wei{\ss}} {et~al.}(2009){Wei{\ss}}, {Kov{\'a}cs}, {Coppin}, {Greve},
  {Walter}, {Smail}, {Dunlop}, {Knudsen}, {Alexander}, {Bertoldi}, {Brandt},
  {Chapman}, {Cox}, {Dannerbauer}, {De Breuck}, {Gawiser}, {Ivison}, {Lutz},
  {Menten}, {Koekemoer}, {Kreysa}, {Kurczynski}, {Rix}, {Schinnerer}, \& {van
  der Werf}}]{Weis:2009qy}
{Wei{\ss}}, A., {Kov{\'a}cs}, A., {Coppin}, K., {et~al.} 2009,
  \href{http://dx.doi.org/10.1088/0004-637X/707/2/1201}{\JournalTitle{\apj},
  707, 1201}

\bibitem[{{Younger} {et~al.}(2008){Younger}, {Fazio}, {Wilner}, {Ashby},
  {Blundell}, {Gurwell}, {Huang}, {Iono}, {Peck}, {Petitpas}, {Scott},
  {Wilson}, \& {Yun}}]{Younger:2008rt}
{Younger}, J.~D., {Fazio}, G.~G., {Wilner}, D.~J., {et~al.} 2008,
  \href{http://dx.doi.org/10.1086/591931}{\JournalTitle{\apj}, 688, 59}

\end{thebibliography}

\end{document}